\documentclass[10pt]{article}

\usepackage[paperwidth=21.0cm,paperheight=29.7cm,textwidth=16cm,textheight=20cm,top=2cm,left=3.cm,bottom=2cm]{geometry}

\usepackage{graphicx,color}
\usepackage{amssymb,amsbsy,amsmath,amsfonts}
\usepackage[mediumspace,mediumqspace,squaren]{SIunits}

\usepackage{fancyhdr}
\usepackage{lineno}
\usepackage{savesym}
\usepackage{amsmath}
\savesymbol{nl}

\definecolor{navy}{rgb}{0,0,0.5}
\usepackage[pdftex,
           hyperindex=true,
           colorlinks=true,
           linkcolor=navy,
           anchorcolor=magenta,
           citecolor=navy,
           urlcolor=navy,
           unicode,
           implicit=true]{hyperref}

\usepackage[ruled,vlined]{algorithm2e}
 
\restoresymbol{TXF}{nl}

\usepackage{url,color,bm,lineno,xspace,soul,ulem,wasysym}
\definecolor{vygreen}{rgb}{0.,0.5,0.3}
\definecolor{vyblue}{rgb}{0,0.3,0.8}
\definecolor{vyred}{rgb}{0.65,0.2,0}
\definecolor{orange}{rgb}{0.8,0.4,0.0}
\definecolor{green}{rgb}{0.2,0.7,0.1}
\renewcommand{\vec}[1]{\boldsymbol{#1}}

\title{\textbf{Modelling Iceberg Capsize in the Open Ocean}}

\author{P. Bonnet$^{1, 2, 3}$, V.A. Yastrebov$^3$, P. Queutey$^4$, A. Leroyer$^4$, A. Mangeney$^{1,5,6}$\\ O. Castelnau$^2$, A. Sergeant$^7$, E. Stutzmann$^1$, J-P Montagner$^1$}
\date{\footnotesize
 {\it $^1$Institut de Physique du Globe de Paris, Sorbonne Paris Cit\'{e}, Seismology Group, CNRS UMR 7154, Paris, France}\\
 {\it $^2$ENSAM, CNAM, Laboratoire Proc\'{e}d\'{e}s et Ing\'{e}nierie en M\'{e}canique et Mat\'{e}riaux, CNRS, Paris, France}\\
 {\it $^3$MINES ParisTech, PSL University, Centre des Mat\'{e}riaux, CNRS UMR 7633, Evry, France}\\
 {\it $^4$Laboratoire en Hydrodynamique Energ\'{e}tique et Environnement Atmosph\'{e}rique (LHEEA),\\ METHRIC Team, CNRS UMR 6598, Centrale Nantes, France}\\
 {\it $^5$Universit\'{e} de Paris, France}\\
 {\it $^6$Inria, Lab. J.-L. Lions, ANGE team, CNRS, France}\\
 {\it $^7$Aix-Marseille Univ, CNRS, Centrale Marseille, LMA UMR 7031, Marseille, France}\\
  }

\begin{document}

\maketitle
%
%

\begin{quote}

\begin{center}
\noindent{\bf Summary} 
\end{center}
At near-grounded glacier termini, calving can lead to the capsize of kilometer-scale unstable icebergs. The transient contact force applied by the capsizing iceberg on the glacier front generates seismic waves that propagate over teleseismic distances. The inversion of this seismic signal is of great interest to get insight into actual and past capsize dynamics. However, the iceberg size, which is of interest for geophysical and climatic studies, cannot be recovered from the seismic amplitude alone. This is because the capsize is a complex process involving interactions between the iceberg, the glacier and the surrounding water. This paper presents a first step towards the construction of a complete model, and is focused on the capsize in the open ocean without glacier front nor ice-m\'{e}lange. 
The capsize dynamics of an iceberg in the open ocean is captured by computational fluid dynamics (CFD) simulations, which allows assessing the complexity of the fluid motion around a capsizing iceberg and how far the ocean is affected by iceberg rotation.
Expressing the results in terms of appropriate dimensionless variables, we show that laboratory scale and field scale capsizes can be directly compared. The capsize dynamics is found to be highly sensitive to the iceberg aspect ratio and to the water and ice densities. 
However, dealing at the same time with the fluid dynamics and the contact between the iceberg and the deformable glacier front requires highly complex coupling that often goes beyond actual capabilities of fluid-structure interaction softwares. Therefore, we developed a semi-analytical simplified fluid-structure model (SAFIM) that can be implemented in solid mechanics computations dealing with contact dynamics of deformable solids. This model accounts for hydrodynamic forces through calibrated drag and added-mass effects, and is calibrated against the reference CFD simulations. We show that SAFIM significantly improves the accuracy of the iceberg motion compared with 
existing simplified models. Various types of drag forces are discussed. The one that provides the best results is an integrated pressure-drag proportional to the square of the normal local velocity at the iceberg's surface, 
with the drag coefficient depending linearly on the iceberg's aspect ratio. 
A new formulation based on simplified added-masses or computed added-mass proposed in the literature, is also
discussed.
We study in particular the change of hydrodynamic-induced forces and moments acting on the capsizing iceberg.
The error of the simulated horizontal force ranges between $5\%$ and $25\%$ for different aspect ratios. 
The added-masses affect the initiation period of the capsize, the duration of the whole capsize being better simulated when added-masses are accounted for.
The drag force mainly affects the amplitude of the fluid forces and this amplitude is best predicted without added-masses.
\end{quote}


\section{Introduction}
\label{sec:intro}
A current concern in climate science is the estimation of the mass balances of glaciers and ice sheets. The Greenland ice sheet mass balance contributes significantly to sea level rise, accounting for about 15\% of the annual global sea level rise between 2003 and 2007 \cite{Zwally2011}. However, it is difficult to draw conclusions on general trends given the high uncertainties \cite{Lemke2013} in these estimations, notably due to difficulties in estimating and partitioning the ice-sheet mass losses \cite{VanDenBroeke2009}.
Ice mass balance can be determined by calculating the difference between the (i) surface mass balance, mainly determined by inland ice gains minus ice losses and (ii) dynamic ice discharge,  mainly made up of submarine melting and iceberg calving \cite{VanDenBroeke2016}.
One third to one half of the ice mass losses of the Greenland ice sheet are due to dynamic ice discharge \cite{Enderlin2014}. Note that dynamic ice discharge is a complex phenomenon, influenced by ocean and atmospheric forcing and by glacier geometry and dynamics \cite{Benn2017}.

When a marine glacier terminus approaches a near-grounded position, calving typically occurs through the capsize of glacier-thickness icebergs. Such buoyancy-driven capsize occurs for icebergs with a width-to-height ratio below a critical value \cite{MacAYEAL2003}.
When drifting in the open ocean, icebergs deteriorate through various processes such as break-up, wave erosion and solar or submarine convection melting \cite{Job1978, Savage2007}, and release freshwater that can potentially affect overturning ocean circulation \cite{Vizcaino2015,Marsh2015}. \cite{Wagner2017} explain that icebergs mainly melt through wind-driven wave erosion that leads to lateral thinning and thus eventually a buoyancy driven capsize of the icebergs. Moreover, iceberg drift simulations have shown that capsizing icebergs live longer and drift further than non-capsizing icebergs  \cite{Wagner2017} .

When they capsize right after calving, icebergs exert an almost horizontal transient contact force on the glacier front. This force is responsible for the generation of up to magnitude $M_\text{w}\approx5$ earthquakes that are recorded globally \cite{Ekstrom2003,Podolskiy2016,Aster2017} and can be recovered from seismic waveform inversion \cite{Walter2012,Sergeant2016}. The study of iceberg calving and capsizing with such glacial earthquakes data is a promising tool, complementary to satellite imagery or airborne optical and radar sensors as it can provide more insights into the physical calving processes and iceberg-glacier-ocean interaction \cite{Tsai2008,Winberry2020} as well as it can track ice losses almost in real time.
However, there is no direct link between the size of an iceberg and the generated seismic signal \cite{Tsai2008, Walter2012,Amundson2012,Sergeant2016,Sergeant2018}. \cite[fig.~6d]{Sergeant2018} showed that a given centroid single-force (CSF) amplitude, which is usually used to model the relevant signal, can be obtained with icebergs of different volumes (their fig. 6d). \cite[fig.~9]{Olsen2019} found a weak correlation between the seismic data (CSF amplitude) and the dimensions of the iceberg, the authors also suggested that taking into account hydrodynamics and iceberg shape could improve this correlation.
Different processes involving the interactions between the iceberg, glacier, bedrock, water and ice m\'{e}lange contribute to the type of calving, earthquake magnitude and seismic waveform \cite{Amundson2010, Tsai2008, Amundson2012}.
 
To investigate in detail the link between iceberg volume, contact force and the generated seismic signals, the use of a hydrodynamic model coupled with a dynamic solid mechanics model is required.
The iceberg-ocean interaction governs the iceberg capsize dynamics and thus controls the time-evolution of the contact force which is responsible for the seismic waveform and amplitudes.
Full modelling of the glacier\,/\,ocean\,/\,bedrock\,/\,iceberg\,/\,ice-m\'{e}lange system is beyond capabilities of most existing models because it requires complex and costly coupling between solid mechanics, contact dynamics and fluid dynamics.
Simplified models of capsizing icebergs proposed in the literature approximate icebergs by 2D rectangular rigid solids subject to gravity and buoyancy force, iceberg-glacier contact force and simplified hydrodynamic effects using either added-masses \cite{Tsai2008} and/or pressure drag \cite{Burton2012,Amundson2012,Sergeant2018,Sergeant2019}.
These models have been proposed to describe a specific aspect of the capsize: its vertical and rotational motion \cite{Burton2012} validated against laboratory experiments, or the horizontal force that icebergs exert on the glacier fronts \cite{Tsai2008,Sergeant2018}.
To build a complete catalogue of seismogenic calving events that can be used for seismic inversion and iceberg characterization, the model must accurately describe
the interaction between the iceberg, glacier and the ocean. 
At the same time, its formulation should either remain sufficiently simple to enable fast simulations of numerous events or, alternatively be based on the interpolation of the response surface constructed on numerous full-model simulations. In particular, the horizontal force and the torque exerted by the fluid on the iceberg should be modelled as accurately as possible, since it controls the horizontal contact force \cite{Tsai2008,Burton2012, Amundson2012, Sergeant2018}.

The present paper aims (i) to provide insights in the complex interactions between a capsizing unconstrained iceberg and the surrounding water in 2D using a reference fluid dynamics solver and (ii) to reproduce the main features of this interaction using a simplified model formulation suitable for being integrated in a more complete model. For this, we use a Computational Fluid Dynamics solver (ISIS-CFD Software for Numerical Simulations of Incompressible Turbulent Free-Surface Flows) to generate reference results for the capsizing motion.
This model solves Reynolds Averaged Navier-Stokes Equations (RANSE) and handles interactions between rigid solids and fluids with a free surface, but is not yet validated for modelling contacts between solids.
This state-of-the-art solver has been extensively validated on various marine engineering cases \cite{Visonneau2005,Queutey2014,Visonneau2016} but not yet applied to kilometre-size objects subject to fast and big rotations like capsizing icebergs in the open ocean, which give rise to high vorticity.
Before applying ISIS-CFD to the field-scale iceberg capsize, we evaluate how well it can reproduce small-scale laboratory experiments (typical dimension of $10\;\centi\meter$).
We compare here ISIS-CFD simulations to the laboratory experiments conducted by \cite{Burton2012}.

To obtain a model that can be easily coupled with a solid mechanics model, we propose a simplified formulation (SAFIM, for Semi-Analytical Floating Iceberg Model) for the interactions between iceberg and water.
In this model, the introduced hydrodynamic forces account for water drag and added-masses, these two effects being considered uncoupled and complementary.
Such a description was initially proposed for modelling the effect of waves on vertical piles \cite{Morison1950} and has been widely used for modelling the effect of waves and currents on bulk structures \cite{Venugopal2009,Tsukrov2002}.
The SAFIM's hydrodynamic forces involve some coefficients that need to be calibrated to represent as accurately as possible the effects of the hydrodynamic flow on the capsize motion. 
These coefficients were calibrated on the reference results provided by ISIS-CFD.

The paper is organized as follows. Section~\ref{sec:ISIS} presents the reference ISIS-CFD fluid dynamics model and its results are compared with those of laboratory experiments from \cite{Burton2012}. The complexity of the fluid motion surrounding the iceberg and the pressure on the iceberg are then discussed. The similarities between the laboratory-scale and field-scale simulations are also presented. 
In Section~\ref{sec:SAFIM}, we present the SAFIM model and discuss the differences with other models from the literature. 
In Section~\ref{sec:Results}, ISIS-CFD and SAFIM are compared for different case-studies, the error quantification and fitting of parameters are discussed. 
Section~\ref{sec:disc} is an overall discussion: comparison of previous models, the new SAFIM model and the reference ISIS-CFD model, followed by a discussion of different drag and added-mass models, which is concluded by a sensitivity analysis with respect to physical properties of geophysical systems.

\section{CFD simulations of iceberg capsize}
\label{sec:ISIS}

\subsection{Problem set-up}
\label{sec:geom}
\begin{figure}[htb!]
\centering
\begin{minipage}{0.6\linewidth}
\centering
\includegraphics[width=1\linewidth]{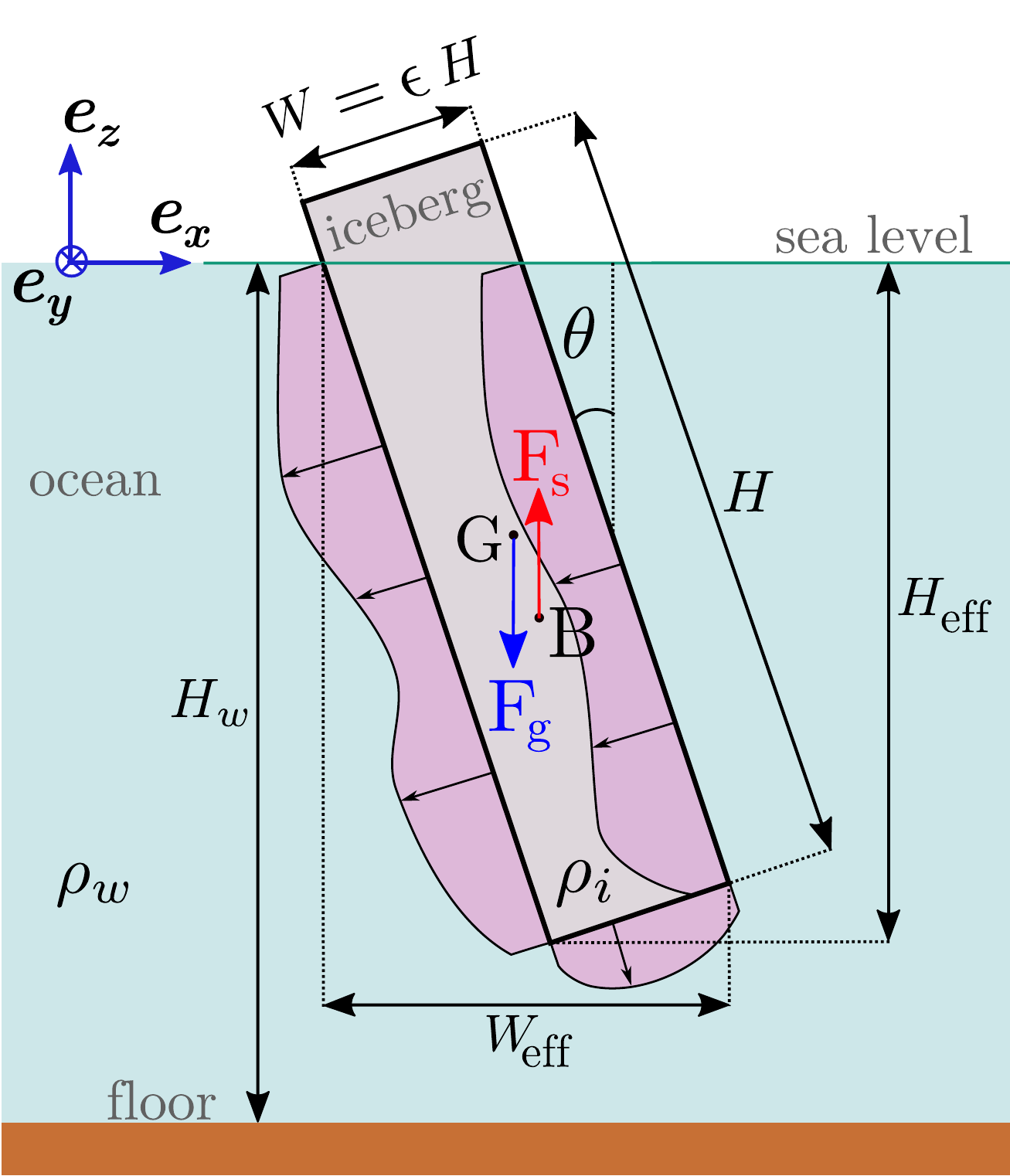}
\end{minipage}
\caption{Simplified iceberg's geometry: centre of gravity is $G$ and the centre of buoyancy is $B$. 
The forces acting on the moving iceberg are: the gravity force $\vec F_g$, the hydrostatic fluid force $\vec F_s$ and the hydrodynamic pressure sketched by the pink shaded area, the forces from the added mass is not indicated; these efforts also induce a torque on the iceberg. \label{fig:schema_iceberg_dz0} 
}
\end{figure}

In this paper, we investigate the capsize of unstable 2D icebergs in the open ocean, i.e. away from the glacier front, other icebergs, and in absence of ice-m\'elange.
Water and ice densities are noted $\rho_w$ and $\rho_i$. In our numerical simulations,
for the field scale, we take the typical values of $\rho_w=1025\;\kilo\gram\per\meter^{3}$ and $\rho_i=917\;\kilo\gram\per\meter^{3}$. For the laboratory experiments, the densities are set to $\rho_w=997\;\kilo\gram\per\meter^{3}$ and $\rho_i=920\;\kilo\gram\per\meter^{3}$, which would enable a direct comparison with~\cite{Burton2012}.
A typical geometry is shown in Fig.~\ref{fig:schema_iceberg_dz0}.
The out-of-plane dimension $L$ of the iceberg (\textit{i.e.} along  $\mathbf{e}_y$) is assumed to be sufficiently large compared to the height and the width such that the problem can be considered essentially two-dimensional. This assumption is discussed in Section~\ref{sec:3d}.

In this two-dimensional set-up, icebergs are assumed to be rectangular with in-plane dimensions $H$ and $W$ and an aspect ratio denoted by $\varepsilon=W/H$.
Rectangular icebergs in a vertical position are unstable, \textit{i.e.} will capsize spontaneously, for aspect ratios smaller than a critical value \cite{MacAYEAL2003}: 
\[
\varepsilon < \varepsilon_c=\sqrt{6\,\rho_i\,\frac{\rho_w-\rho_i}{\rho_w^2}}. 
\]
This critical aspect ratio is $\varepsilon_c\approx0.75$ for the field densities and $\varepsilon_c\approx0.65$ for the laboratory densities.
For $\varepsilon > \varepsilon_c$, icebergs are vertically stable and will not capsize unless initially tilted sufficiently \cite{Burton2012}.

The iceberg is assumed to be homogeneous and rigid, \textit{i.e.} it does not deform elastically.
The mass of an iceberg per unit of thickness along $\mathbf{e}_y$ is given by $m = \rho_i H^2 \varepsilon$.
Points $G$ and $B$ are the centre of gravity of the iceberg and the centre of buoyancy, respectively.
The iceberg position is described by the horizontal and vertical positions of $G$, denoted $x_G$ and $z_G$, respectively, 
and by the inclination $\theta$ with respect to the vertical axis, which is collinear with the vector of the gravity acceleration. 
$H_w$ is the water depth.

\subsection{ISIS-CFD solver}

The ISIS-CFD solver, developed by LHEEA in Nantes (France), is a state-of-the-art solver for the dynamics of multiphase turbulent flows \cite{Queutey2007,Leroyer2011,Guilmineau2017,Guilmineau2018}, interacting with solid and/or flexible bodies \cite{Hay2006, Leroyer2005, Durand2014}, and with a free surface.
Today, it is one of very few available software products capable of solving problems as complex as interactions between solids and fluids with a free surface (water and air interface).
The target applications of ISIS-CFD are in the field of marine engineering, e.g. modelling the dynamic interactions between a ship and surface waves \cite{Visonneau2005, Queutey2014, Visonneau2016} or the complex flows and interactions involved in the global hull-oars-rower system in Olympic rowing \cite{Leroyer2012}.
ISIS-CFD solves the Reynolds Averaged Navier-Stokes Equations (RANSE) \cite{Robert2018} and also disposes few other turbulence models.

\begin{figure}[htb!]
\centering
\includegraphics[width=1\linewidth]{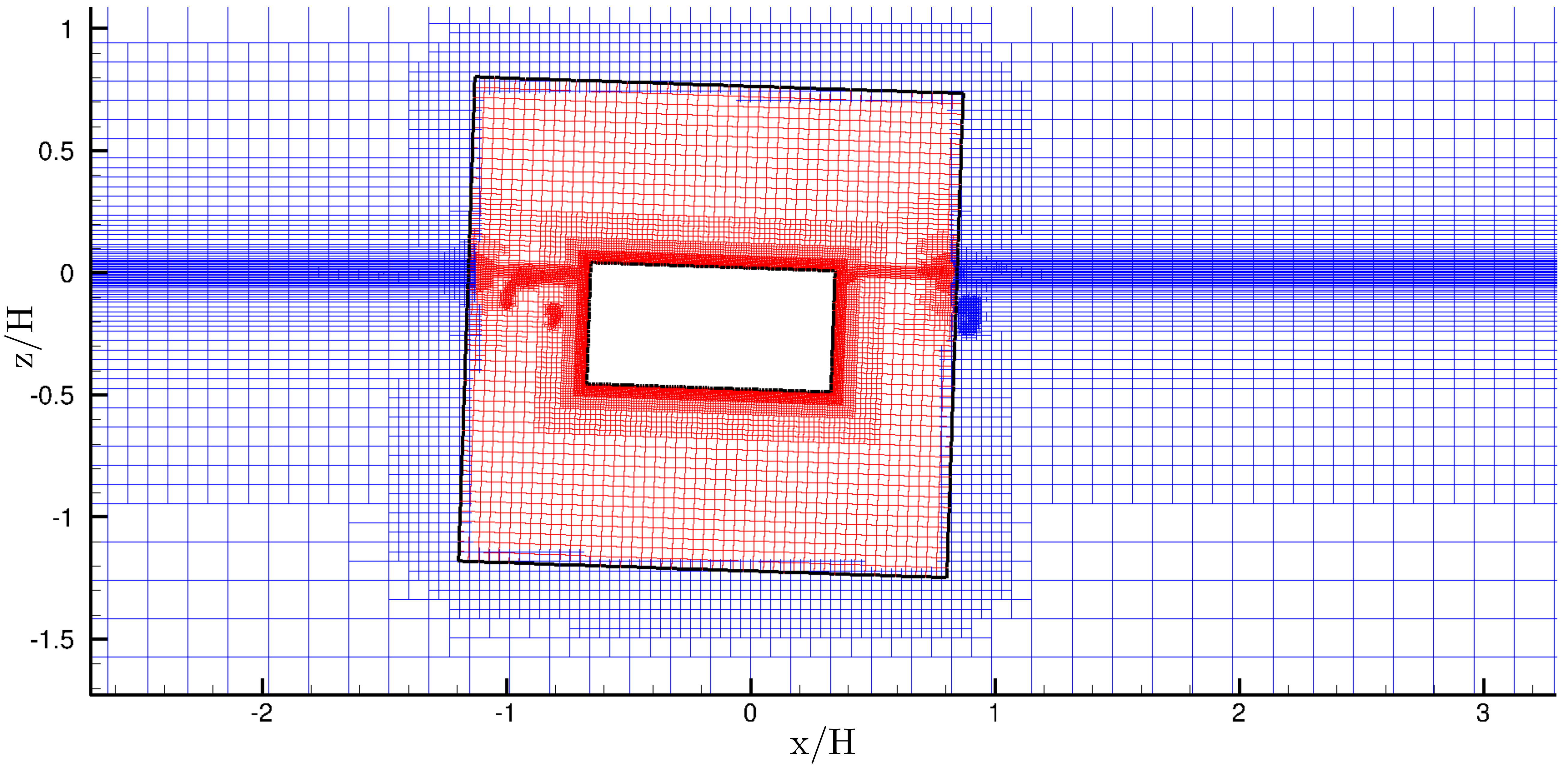}
\caption{A portion of a typical mesh used for the simulation of iceberg capsize with ISIS-CFD. 
The two axes represent the dimensionless horizontal position $x/H$ and the dimensionless vertical position $z/H$. The iceberg $\varepsilon=0.5$ (in white) is in the centre of a squared domain (fine mesh in red) moving and rotating with the iceberg over a background domain (coarser mesh in blue). The mesh is automatically refined around the air-water interface  (a finely meshed horizontal zone) and also around the solid/fluid interface.}
\label{fig:mesh}
\end{figure}

For the specific application of iceberg capsize (Fig.~\ref{fig:schema_iceberg_dz0}), two different turbulence equations were tested and were found to give very similar results: k-w \cite{Menter1993} and Spalart-Allmaras \cite{Spalart1992}.
The code uses an adaptive grid refinement \cite{Wackers2012} or an overset meshing method (mandatory to deal with large amplitude body motion close to a wall for example) to connect two non-conforming meshes. The mesh used here is a converged mesh with $n=43\,000$ elements.
An example of a typical mesh is illustrated in Fig.~\ref{fig:mesh}.
The coupling between the solid and the fluid is stabilized with a relaxation method based on the estimation of the periodically updated added-mass matrix \cite{Yvin2018}. The lateral sides of the simulation box are put far from the capsizing iceberg, and include a damping region, so that the reflected waves do not interfere with the flow near the iceberg for the duration of the simulation.

In the field of application of the ISIS-CFD flow solver, the typical range of Reynolds numbers ($Re$) extends from $10^6$ for model-scale ship flow to $10^9$ for full-scale ship flow \cite{Visonneau2006}, and the local viscous contribution to the hydrodynamic force is as high as $\approx50$ \% of the total drag force.
Here, the ISIS-CFD solver is applied to a significantly different geometry (rectangular shape of iceberg instead of streamlined shape of ship), type of motion (big rotations instead of translational motion) and dimensions (km-scale icebergs instead of tens of meters long ships). 
This application provides new challenges for the ISIS-CFD solver: high vorticity and free surface motion, greater lengths and velocities together with massive separations due to the sharp corners of the iceberg.

\subsection{Laboratory experiments }

\label{sec:Burton}

\begin{figure}[htb!]
\centering
\includegraphics[width=1\linewidth]{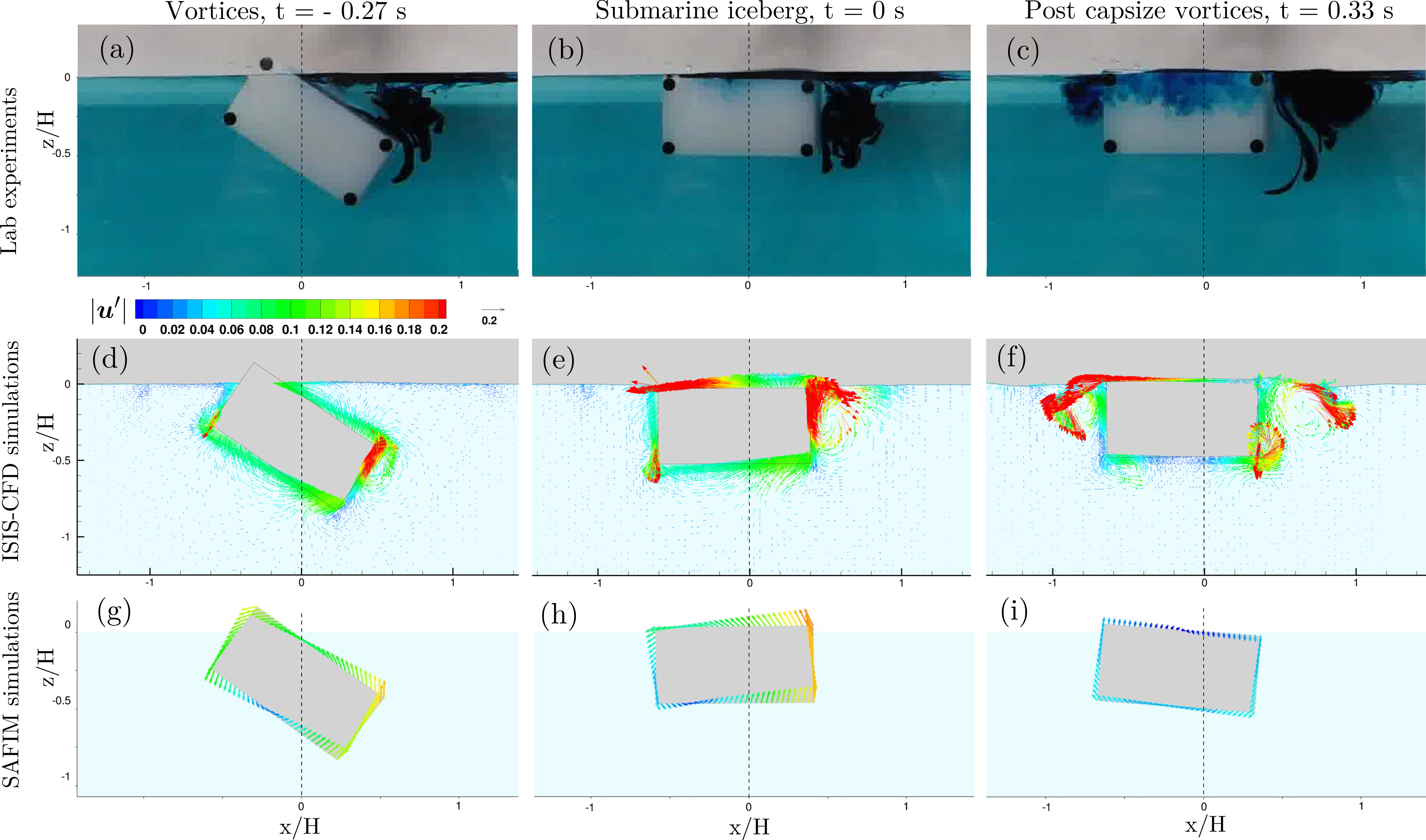}
\caption{Side view of experiments from \cite{Burton2012} (a, b, c); colour map of the dimensionless velocity $\vec u'=\vec u / \sqrt{gH}$, where $\vec u$ is the velocity field of the fluid, in the water computed with ISIS-CFD (d, e, f); dimensionless velocity along the surface of the iceberg with SAFIM (g, h, i). Three time moments of the capsize (indicated on top of each column) are shown. The time scale is calibrated such that $t=0$ s corresponds to the first time when the iceberg reaches $\theta=90\degree$, as in \cite{Burton2012}. The iceberg aspect ratio is $\varepsilon=0.496$. The floor and walls are not shown in the snapshots. The corresponding animation of ISIS-CFD simulation in available in supplementary material [S1].
}
\label{fig:comp}
\end{figure}

Since ISIS-CFD simulations is compared with laboratory data (Section~\ref{sec:complab}), 
we briefly summarize here the technical details of the corresponding experiments conducted by \cite{Burton2012}.
The laboratory experiments consist in the capsize of parallelepipedic plastic iceberg proxies of density $\rho_i=920 \pm 1\; \kilo\gram\per\meter^{3}$ in a long and narrow fjord-like tank $244~\centi\meter$ long, $28~\centi\meter$ wide and $30~\centi\meter$ tall, filled with water at room temperature ($\rho_w\approx997\; \kilo\gram\per\meter^{3}$). 
To assess the effect of water depth $H_w$ on iceberg-capsize dynamics, two types of experiments were conducted in which the water height was varied from $11.4\;\centi\meter$ to $24.3 \;\centi\meter$.
The iceberg height was $H=10.3 \;\centi\meter$, ($H<H_w$) and the width varied between $W=2.5 \;\centi\meter$ and $W=10.2 \;\centi\meter$, corresponding to aspect ratios ranging between $0.25$ and $1$. The length of the iceberg was $L=26.6 \;\centi\meter$, which is slightly smaller than the tank width to reduce edge effects so that the flow can be considered as two-dimensional.
The plastic icebergs were initially placed slightly tilted with respect to the vertical position and were held by hand close to hydrostatic equilibrium.
When the surface of the water became still, several drops of dye were introduced around the plastic iceberg to visualize the water flow. Then the icebergs were released to capsize freely.
The capsizes were recorded by a camera located outside the tank. Snapshots are shown in Fig.~\ref{fig:comp} (top row).
Further experimental details can be found in \cite{Burton2012}.
A selection of four experiments are presented here, corresponding to aspect ratios $\varepsilon=0.246$, $\varepsilon=0.374$, $\varepsilon=0.496$ and $\varepsilon=0.639$.

Laboratory experiments show, to some extent, the fluid motion using dye at some specific locations. The ISIS-CFD computational fluid dynamics model makes a valuable contribution to the understanding of the complex motion of the fluid surrounding a capsizing iceberg since it computes the whole velocity field in the fluid.
Fluid velocity colour maps computed with ISIS-CFD are qualitatively compared with the images of the laboratory experiments in Fig.~\ref{fig:comp} (a-f), the maps of the iceberg's surface velocity computed using the calibrated SAFIM model (see Section~\ref{sec:SAFIM}) are also shown in Fig.~\ref{fig:comp} (g,h,i).  All results are shown at the identical time moments centred at the time when $\theta=90\degree$. 

Results for the capsize of an iceberg of aspect ratio $\varepsilon=0.496$ are shown for three different times: in Figs.~\ref{fig:comp}~(a,d,g) during capsize; in Figs.~\ref{fig:comp}~(b,e,h) when the iceberg reaches the horizontal position for the first time ($\theta=90\degree$); and in Figs.~\ref{fig:comp}~(c,f,i) some time later. 
The arrows represent the dimensionless velocity $|u'|=|\vec u| / \sqrt{gH}$, where $\vec u$ is the velocity field of the fluid.
We observe a good qualitative agreement between the position and inclination of the iceberg obtained by ISIS-CFD and the laboratory experiments.
Note that the iceberg is submarine when it reaches $\theta\approx90\degree$ for the first time (Figs.~\ref{fig:comp} (b,e)).
The motion of the fluid - initially almost at rest - is visible all around the capsizing iceberg.
Large vortices, associated with the iceberg motion, are clearly visible throughout the capsize in Fig.~\ref{fig:comp} (top and middle row).
The intense fluid motion represents an important amount of kinetic energy that is eventually dissipated; this energy is transmitted by the motion of the iceberg, and this slows down the iceberg.
 Note also that the iceberg moves leftwards during the capsize.

\subsection{Comparison of simulations with experiments \label{sec:complab}}

\begin{figure}[htb!]
  \includegraphics[width=1.\linewidth]{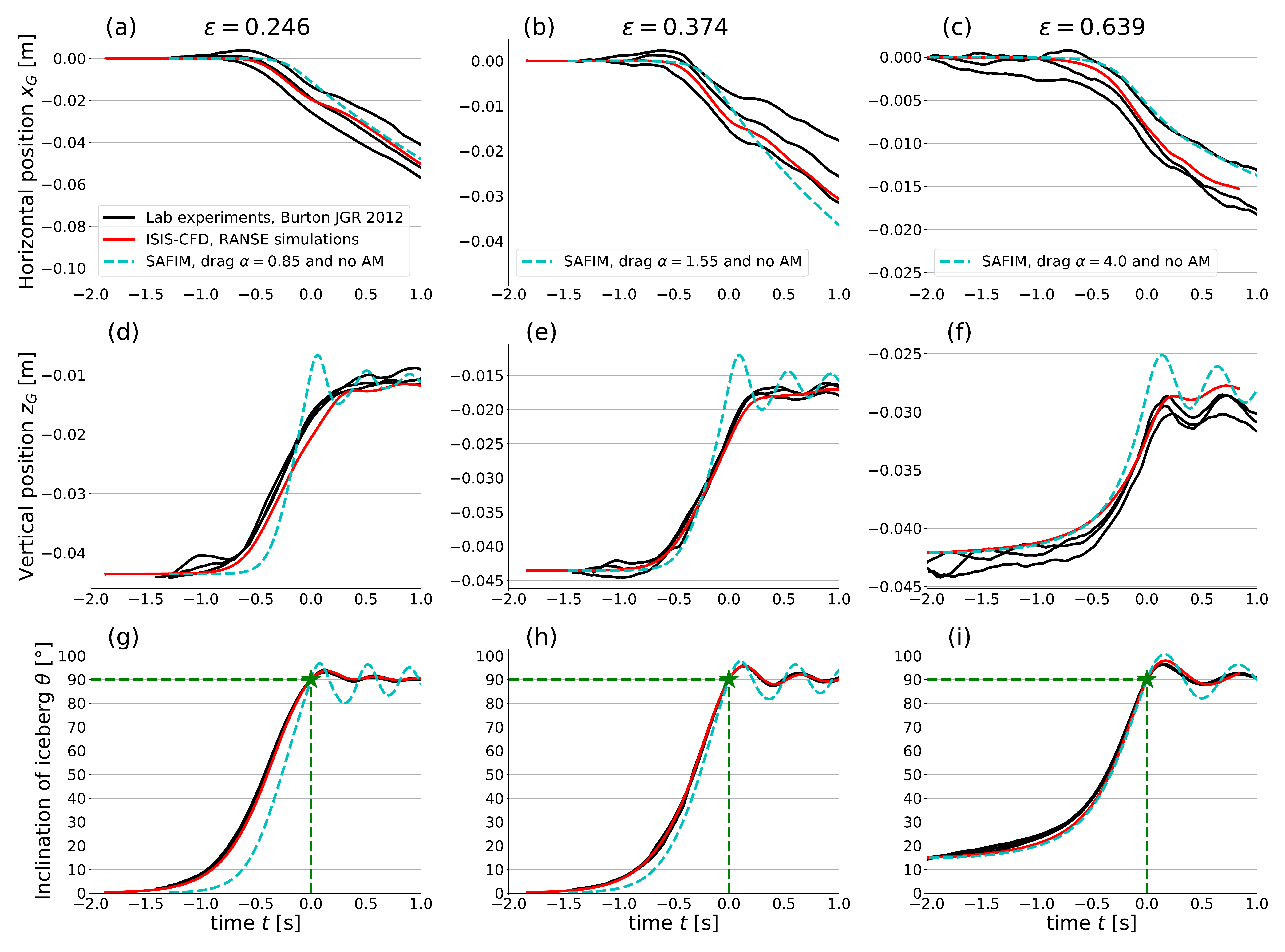}
    \caption{Horizontal position $x_G$ (a, b, c), vertical position $z_G$ (d, e, f) and tilt angle $\theta$ (g, h, i) of icebergs of height $H=10.3$ cm for $\varepsilon=0.246$ (a, d, g), $\varepsilon=0.374$ (b, e, h) and $\varepsilon=0.639$ (c, f, i). Data are provided for three laboratory experiments (solid black), ISIS-CFD simulations (solid red) and SAFIM simulations (dashed blue). The origin of the time axis in all experiments and simulations is set to the time at which the iceberg reaches the horizontal position for the first time, i.e. when $\theta=90\degree$ (green stars and dashed lines), as in \cite{Burton2012}}
  \label{fig:BurtonISISSAFIM}
\end{figure}

Here, we compare ISIS-CFD with the laboratory experiments by \cite{Burton2012}.
In Fig.~\ref{fig:BurtonISISSAFIM}, results are provided for three different unstable icebergs: 
a thin $\varepsilon~=~0.246$ iceberg, 
a medium $\varepsilon~=~0.374$ iceberg 
and a thick $\varepsilon~=~0.639$ iceberg.
The horizontal position $x_G$, vertical position $z_G$ and tilt angle $\theta$ are plotted against time.
As the plastic icebergs were initially positioned by hand, some variability in the results are observed.
To provide an estimate of the variability of the protocol, three experiments with identical plastic icebergs and the same (nominal) initial conditions were conducted for each aspect ratio.
We selected these three aspect ratios because of the consistency of the experimental results.
The initial conditions in ISIS-CFD were chosen to fit the average values of the laboratory experiments. 
In ISIS-CFD simulations, the icebergs were tilted by a small angle of $0.5\degree$ for the thin and medium iceberg 
(black curves in Figs.~\ref{fig:BurtonISISSAFIM}(d,e)) and a larger angle of $15\degree$ for the thicker iceberg (black curves in Fig.~\ref{fig:BurtonISISSAFIM}(i)). 
The icebergs were initially placed in a hydrostatic equilibrium.
The water level in the tank $H_w=11.4\;\centi\meter$ or $H_w=24.3\;\centi\meter$ was found to have a negligible effect on the iceberg motion: results are within the data variability shown in Fig.~\ref{fig:BurtonISISSAFIM}.
Therefore, the experiments with a constant water depth $H_w=24.3~\centi\meter$ are compared with the ISIS-CFD simulations carried out for the same water depth.

We first analyse the motion of the iceberg during capsize. Once released, it tilts to reach the horizontal position with associated upward and sideward motion.
It then rises out of the water in a rocking motion superimposed with a continuous horizontal displacement (Fig.~\ref{fig:BurtonISISSAFIM}). 
The thinner the iceberg, the faster it moves in the horizontal direction with a quasi constant velocity at least for the first $1.5$~s.
This horizontal motion is an important aspect of the iceberg capsize on which we would like to focus here.
Note that, besides gravity and buoyancy which cannot cause horizontal motion, the only external force acting on the capsizing iceberg is due to the relative motion of water around the iceberg (the air has a negligible effect here).
These hydrodynamic forces are responsible for the horizontal iceberg motion.
They need to be captured accurately by the model as they contribute considerably to the contact force generated between the iceberg and the glacier front when a just-calved iceberg capsizes \cite{Tsai2008,Sergeant2018}.

Fig.~\ref{fig:BurtonISISSAFIM} shows that ISIS-CFD results are in very good agreement with the laboratory data especially in terms of the evolution of the tilt angle.
The slight discrepancy on the vertical and rotational motion computed by ISIS-CFD could be due to differences between the laboratory and simulation set-ups with regards to the 2-dimensional approximation and the initial conditions as discussed above.
Another reason for this slight discrepancy could be related to the turbulence model treated by the RANS approach. The generation of large vortices and separations are not initially induced by turbulent phenomena. We observed that Euler approach (perfect fluid with no viscosity and thus no possible dissipation of energy in turbulence phenomenon) 
captures similar flow topologies. However, the evolution of these vortices and separations can be affected by turbulent effects for which the RANS approach is not specifically designed for. Simulations using methods such as DES (detached eddy simulation) or LES (large eddy simulation) could improve the accuracy but would require high computational costs.

\subsection{From the laboratory to the field scale}
\label{sec:lab_scale}
  
In the previous section, ISIS-CFD simulations were shown to fit laboratory experiments very well.
However, our aim is to reproduce the dynamics of the capsize of field-scale icebergs with dimensions of several hundred metres,\textit{ i.e.} four orders of magnitude larger than for the laboratory scale.
Also, as pointed out by \cite{Sergeant2018}, the laboratory-scale Reynolds number $Re=LU/\nu \approx10^3$, is six orders of magnitude smaller than the characteristic Reynolds number $Re\approx 10^{9}$ for the field scale (with $L$ the typical length, $U$ the typical speed and $\nu$ the dynamic viscosity of the fluid). Global viscous effects are expected to be more pronounced for laboratory-scale than for the field-scale capsize.
Therefore, the question is whether laboratory-scale experiments can be used to understand the kinematics of the field-scale iceberg capsize.

\begin{figure}[htb!]
\centering
\includegraphics[width=1\linewidth]{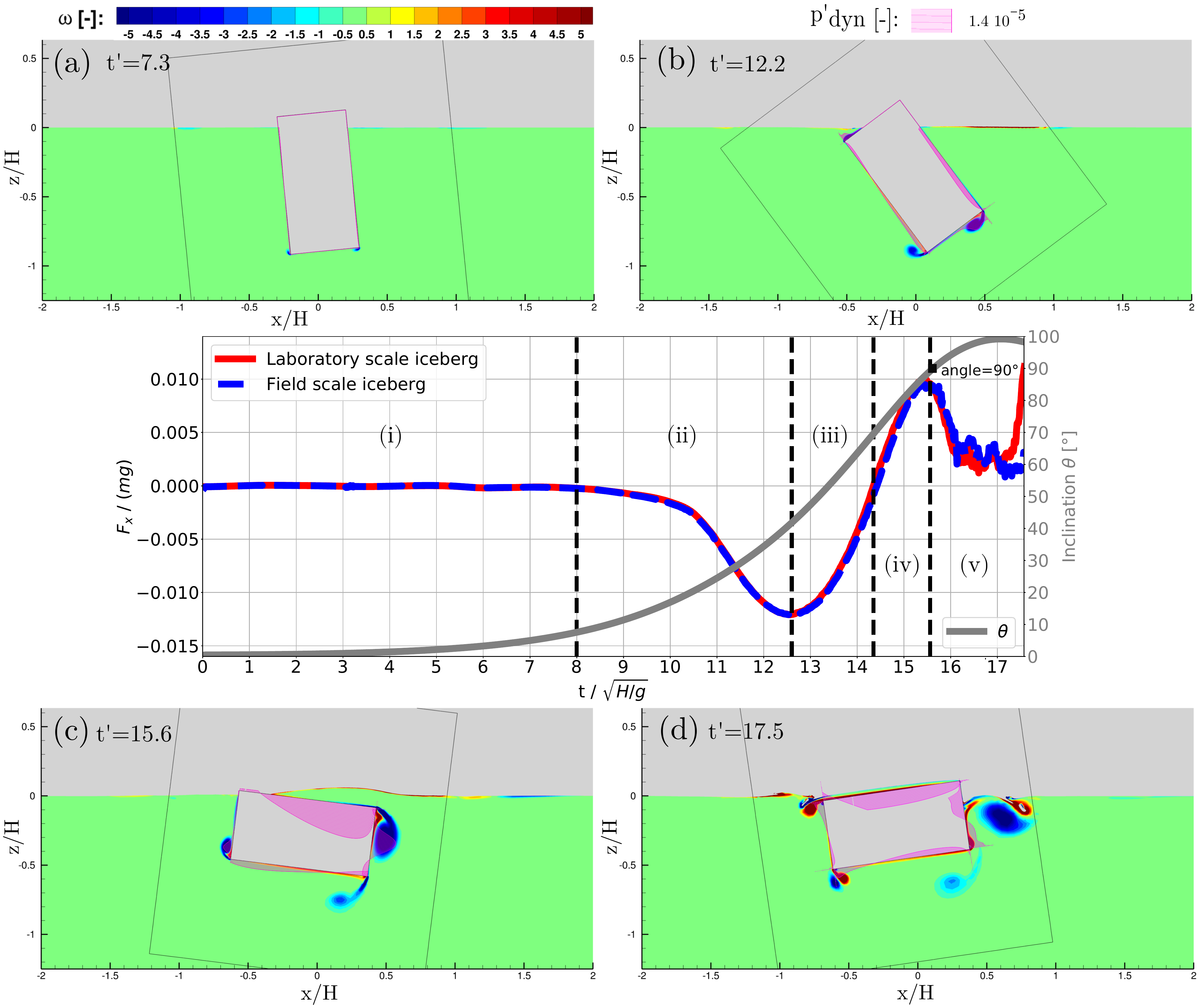}
\caption{The horizontal dimensionless force $F'=F_x/(mg)$ that the water exerts on the iceberg ($\varepsilon = 0.5$) is plotted against a dimensionless time $t'=t/T_H$ with the characteristic time $T_H=\sqrt{H/g}$ for both laboratory ($H=0.103~\meter$, $T_H=0.102$ s, in red) and field ($H=800~\meter$, $T_H=9.03$ s, in blue) scales.
The iceberg inclination $\theta$ (grey curve with scale on the right) is plotted against dimensionless time $t'$, the curve is the same for the lab scale and the field scale. 
The vorticity fields around the iceberg at four different times are also shown: (a) initial phase of iceberg's motion $t'\approx7.3$, (b) time of maximal force in the left direction  $t'\approx12.2$, (c) time of maximal force in the right direction $t'\approx15.6$ and (d) corresponds to the oscillations of the iceberg at later time $t'\approx17.1$. The pink shaded area represents the local hydrodynamic pressure. The colour maps are discussed in Section~\ref{sec:pressureloc} and the phases (i) to (v) are discussed in \ref{sec:lab_scale}.}
\label{fig:Fx_T1_L1_ISIS}
\end{figure}

We compare the horizontal force induced by the water to the iceberg during its capsize computed by ISIS-CFD for the two cases: 
(1) a field-scale iceberg of height $H=800\;\meter$ and 
(2) a laboratory-scale iceberg of height $H=0.103\;\meter$, 
all other parameters being the same: the aspect ratio $\varepsilon=0.25$, 
infinite water pool, 
same densities of the water and the ice, taken to be equal to the field densities (Section~\ref{sec:geom}).
Results are given in Fig.~\ref{fig:Fx_T1_L1_ISIS} using dimensionless variables,\textit{ i.e.} a dimensionless horizontal force  $F'_x=F_x/(mg)$ acting on the capsizing iceberg and a dimensionless time $t'=t/\sqrt{H/g}$. Note that this horizontal force $F'_x$  acting on the iceberg is the hydrodynamic force. However, in the case of iceberg-glacier interaction a similar hydrodynamic force will contribute to the total contact force between a glacier and a capsizing iceberg.
We observed that the two curves corresponding to the two scales are very similar from the beginning of the movement until $t'\approx15.6$, which corresponds to $\theta\approx 90\degree$.
This similarity between the forces at the laboratory and field scales can be explained using the Vaschy-Buckingham $\pi$ theorem, assuming that the effect of viscosity is negligible, as detailed in Appendix~\ref{sec:simi}.
For times larger than $t' \approx 15.3$, the discrepancy between laboratory and field scales is larger and dimensions start to play a more important role.
This discrepancy probably originates from the fact that after the buoyancy driven capsize, the iceberg motion is driven by the evolution of complex vortices and different gravity-waves dynamics.

The other variables of the system (vertical force and torque acting on the iceberg and horizontal and vertical displacement and inclination of the iceberg) are also similar for the laboratory and field scales.

Since it was demonstrated that the laboratory and field scales produce the same horizontal dimensionless force, in the remaining simulations we will present only the dimensionless quantities obtained from simulations of the laboratory-scale iceberg with $H=0.103\,\meter$, for densities corresponding to field values, and in absence of sea floor. 
The laboratory scale was chosen because numerical convergence is easier to achieve in ISIS-CFD for the laboratory scale than for the field scale. The sensitivity of the capsize to the densities will be discussed in Section~\ref{sec:density}. Also, the depth of the sea floor was observed to have no significant effect on the capsize dynamics.
Results for icebergs of different heights (but of the same aspect ratio) can be deduced with a factor of proportionality given by the normalizations from Table~\ref{tab:adim}.

\begin{table}
\begin{center}
  \begin{tabular}{ l c}
    Variable name & Dimensionless variable \\ \hline
    forces & $F' = F/(mg)$ \\
    torque &  $M'_{\theta}=M_\theta/(mgH)$ \\
    positions & $r' = r/H$\\
    time & $t' = t\sqrt{g/H}$ \\
    velocity & $u' = u / \sqrt{Hg}$\\
  \end{tabular}
\end{center}
\caption{Table of dimensionless variables, with $N'$ denoting the dimensionless variable related to $N$ and with the iceberg linear mass density $m= \rho_i\,H^2\,\varepsilon$ G the centre of gravity of the iceberg and the characteristic time $T_H=\sqrt{H/g}$. Note that the dimensionless forces and torques can also be written through a normalization by the characteristic mass $m$, length $H$ and time $T_H$ with the following formulas: $F'_x=F_x \,T_H^{2}/ (m\,H)$, $F'_z=F_z \,T_H^{2}/ (m\,H)$ and $M'_{\theta}=M_{\theta} \,T_H^{2}/ (m\,H^2)$. See Section.~\ref{sec:density} for a discussion on a non-dimensionalization with a dimensionless time $T_{\rho,H}$ that depends on the densities.}
\label{tab:adim}
\end{table}

In ISIS-CFD simulations and laboratory experiments, we observe five stages during the capsize:
\begin{enumerate}
\item In the \textit{initial} phase ($0<t'<6$), the horizontal force $F'_x$ oscillates around zero with a negligible amplitude (about $1\%$ of its extremum amplitude). This stage is the initiation of the capsize with buoyancy and gravity forces making the iceberg rotate and rise.
\item Then the absolute value of $F'_x$ increases, first slowly and then faster until the first extremum at $t'\approx12.2$. This is explained by the fact that the induced vertical and rotational velocities and accelerations of the iceberg produce a hydrodynamic force that has a non-zero horizontal component, which is the only horizontal force acting on the iceberg. It induces a horizontal motion of the iceberg towards the left for the anti-clock wise iceberg rotation considered here.
\item The absolute value of $F'_x$ decreases to $F'_x=0$ before $t'\approx14.4$, which corresponds to $\theta\approx70\degree$ (Fig.~\ref{fig:Fx_T1_L1_ISIS}). The horizontal motion of the iceberg triggers a horizontal resisting fluid force.
\item The force $F'_x$ becomes positive and increases to an extremum at $t'\approx15.6$, where the iceberg is horizontal $\theta\approx90\degree$. Its amplitude is of the same order of magnitude than the first negative force but the duration is shorter. Therefore, it decelerates the iceberg leftward motion, but does not cancel it.
\item At the later stage (after $t'\approx15.6$), $F'_x$ oscillates around zero and is slowly damped. The iceberg rocks around $\theta=90\degree$ and drifts to the left while slowly decelerating.
\end{enumerate}

\begin{figure}[htb!]
\centering
\includegraphics[width=1\linewidth]{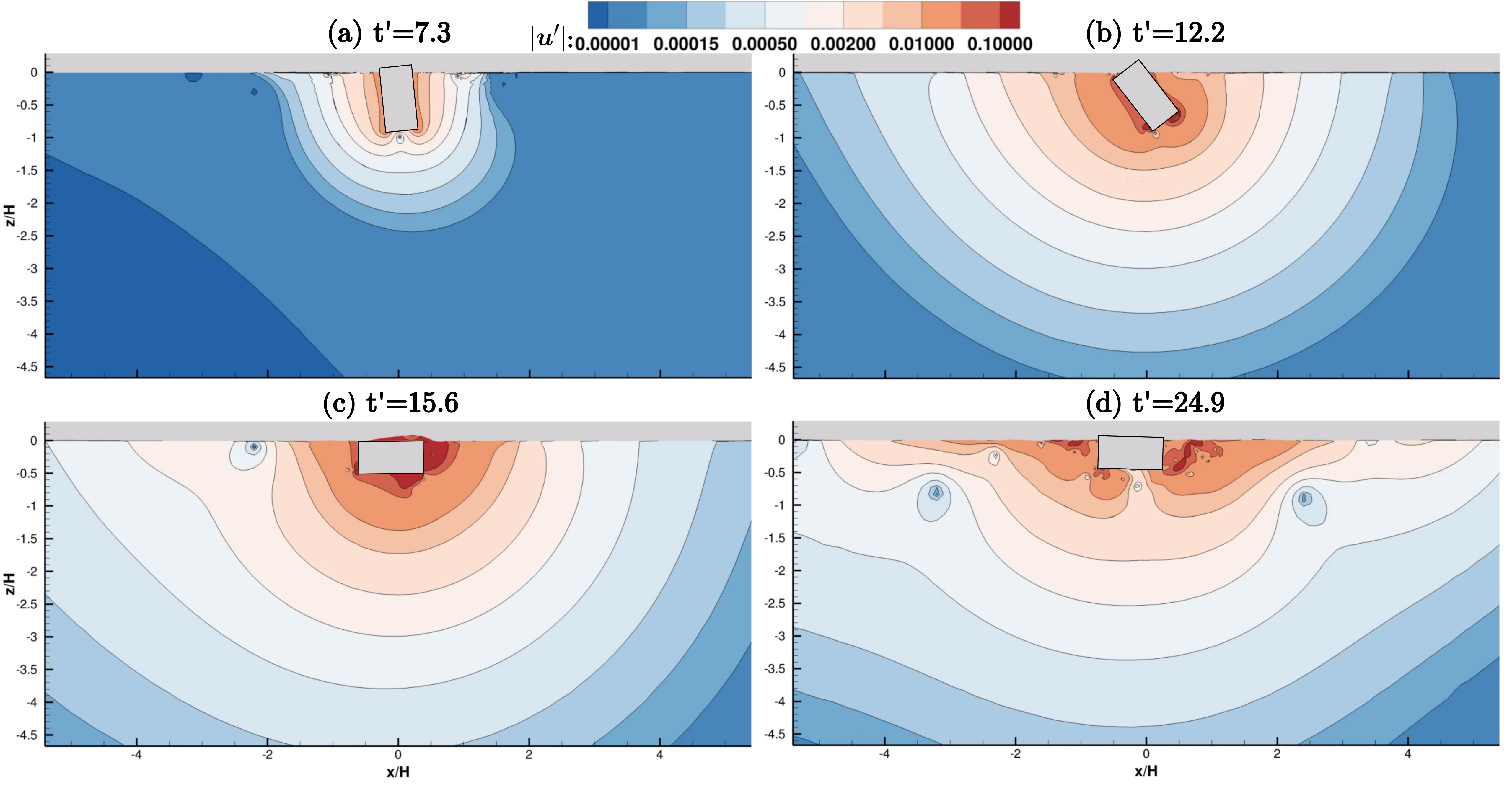}
\caption{Iso-lines of the absolute value of the dimensionless velocity $|u'|=|u / \sqrt{gH}|$ in the fluid surrounding the iceberg with aspect ratio $\varepsilon=0.5$ (grey rectangle). Captions (a), (b) and (c) correspond to the same times as in Fig. \ref{fig:Fx_T1_L1_ISIS} (a-c). Figure (d) at time $t'=24.9$ shows the vanishing fluid motion. An animated figure is available in supplementary material [S2].}
\label{fig:isov}
\end{figure}

The highest water velocities in the surrounding ocean are reached when the iceberg is close to $\theta=90\degree$. Dimensionless velocities are shown in Fig~\ref{fig:isov}. We observe that for an iceberg of height $H$ (here $800\; \meter$) :
\begin{itemize}
\item high velocities in the fluid $\approx0.5\,\sqrt{g\,H}$ (here $\approx 42\; \meter/\second$) are reached between times $t'=10$ (here $\approx 90\; \second$) and $t'=34$  (here $\approx 307\; \second$), see dark red regions in Figs.~\ref{fig:isov}~(b,c,d)).
\item at a distance of about $H$ from the iceberg, the water flows at a maximum speed of $\approx0.01\,\sqrt{g\,H}$ (here $\approx88\; \centi \meter/\second$).
\item at a distance of about $3.5\,H$ from the iceberg, the water flows at a maximum speed of $\approx 0.0005\,\sqrt{g\,H}$ (here $\approx4.4\; \centi \meter/\second$).
\end{itemize}

Note that the maximum plume velocities measured in \cite{Mankoff2016, Jouvet2018} are horizontal velocities at the surface $\approx 3 \;\meter/\second$.

Moreover, we observe that the iso-lines for the velocities are roughly semi-circles centered on the iceberg, with a radius slightly higher in horizontal direction.
 
\section{Empirical model SAFIM}
 \label{sec:SAFIM}

\subsection{General formulation}

The reference ISIS-CFD model has the advantage of being very accurate for fluid-structure interactions but it cannot readily model contacts between deformable solids.
As explained in the introduction, we aim to construct a simpler fluid-structure interaction model that can be more easily coupled with dynamic solid-mechanics models.
Thus we propose a simple empirical model that can be used to estimate the horizontal force applied to a capsizing iceberg.
This model was initiated in \cite{Sergeant2018, Sergeant2019} and is extended and validated in this study.

As proposed by \cite{Tsai2008}, \cite{Burton2012} and \cite{Sergeant2018}, one possible way to construct such a simple model of a capsizing iceberg consists in solving equations of a rigid iceberg motion subject to relevant forces and torque while discarding water motion. The general equations of iceberg motion for such simplified models can be written in two dimensions as:
\begin{eqnarray}
\label{eqn:alleq1}
(m + m_{xx} ) \ddot{x}_G + m_{xz}\ddot{z}_G+J_{x\theta}\ddot{\theta} &=& \vec{F}_d \cdot\vec e_x\\
\label{eqn:alle2}
m_{xz}\ddot{x}_G+(m + m_{zz} ) \ddot{z}_G + J_{z\theta}\ddot{\theta} &=& \left(\vec F_g + \vec F_s + \vec F_d\right)\cdot\vec e_z\\
\label{eqn:alleq3}
J_{x \theta}\ddot{x}_G +J_{z \theta}\ddot{z}_G+ (I + I_{\theta \theta}) \ddot{\theta} &=& \left(\vec M_s + \vec M_d\right)\cdot\vec e_y
\end{eqnarray}
where $I=\rho_i H^4 \varepsilon (1+\varepsilon^2)$ is the moment of inertia of the iceberg with respect to its centre of gravity $G$ and around an axis parallel to $\mathbf{e}_y$ (multiplying $I$ by the iceberg thickness along $\mathbf{e}_y$ gives the inertia for the three-dimensional case).
Such a formulation accounts for the hydrostatic force $\vec F_s$ and the corresponding torque $\vec M_s$ computed at $G$, the gravity force $\vec F_g$, overall hydrodynamic effects expressed by the force $\vec F_d$ and the associated torque $\vec M_d$ at $G$ and so-called added-masses $m_{xx}$, $m_{zz}$, $I_{\theta \theta}$, $m_{xz}$, $J_{z \theta}$ and $J_{x \theta}$ that account for the mass of water that must be accelerated during the iceberg motion.

Hydrodynamic forces that oppose the motion of the iceberg are commonly called \textit{drag forces} $\boldsymbol F_d$. The corresponding drag torque $\boldsymbol M_d$ accounts for a particular distribution of the drag pressure along the iceberg surface.
The drag force usually scales with the squared relative velocity between the water and the solid with a factor of fluid density and it acts in the opposite direction of this velocity. Note that the friction drag can be neglected here, as shown in Appendix~\ref{sec:simi}.

When the water motion is not computed, the added-mass (AM) should also be included in the model.
Added-masses introduce some additional inertia to the moving iceberg.
This effect is known to be significant when the density of the fluid is comparable or bigger than the density of the solid, such as for ice and water.
The matrix of added-masses, which is symmetrical \cite{Yvin2018,Molin2002}, has the following form:
\begin{equation}
[m_{AM}] =
\begin{bmatrix}
m_{xx}       & m_{xz} & J_{x \theta} \\
m_{xz}       & m_{zz} & J_{z \theta} \\
J_{x \theta} & J_{z \theta} & I_{\theta \theta}
\end{bmatrix}
\end{equation}
Added-mass effects are of two types: a force associated with an added-mass can arise in a given direction due to (1) an acceleration in that direction, which corresponds to
the \textit{diagonal} terms $m_{xx}$, $m_{zz}$ and $I_{\theta \theta}$ in Eq.~(\ref{eqn:alleq1}-\ref{eqn:alleq3}), and (2) an acceleration in another direction, which is accounted for by the \textit{coupled} terms $m_{xz}$, $J_{z \theta}$, $J_{x \theta}$.

Within this framework, models proposed by \cite{Burton2012} and \cite{Tsai2008}, summarized in Table~\ref{tab:previous_models}, differ in the way they account for the drag and the added-mass.
In the formulation proposed by \cite{Tsai2008}, pressure drag and the associated torque are not considered and only some of diagonal terms are taken into account in the added-mass matrix.
In the formulation by \cite{Burton2012}, added-mass effects are neglected. As for the drag effects, they are assumed to depend only on individual components of the velocity of $G$ and on the angular velocity, for example the drag along $\vec e_x$ only depends on the velocity $\dot x_G$.
As a consequence, both models predict that an iceberg initially at rest ($\dot x_G=0$) will not experience any horizontal movement along $\vec e_x$ during its capsize.
As discussed above, this result contradicts experimental and ISIS-CFD results.

\begin{table}
\begin{center}
  \begin{tabular}{ l l l }
     Ref. $\displaystyle\vphantom{\frac{A}{A}}$& Added-mass (AM):          & Drag:\\
     \hline    
       \cite{Tsai2008} $\displaystyle\vphantom{\frac{A}{A}}$&  $m_{xx_{T}}=\frac{3 \pi \rho_w L}{8}  \left(H^2\cos(\theta)^2 + W^2 \sin(\theta)^2\right)$    & $\vec F_d = 0 $ \\
          & $m_{zz_{T}} = 0$                       & $\vec M_d = 0$ \\
          & $I_{\theta\theta_{T}}=\frac{\rho_w}{24}(H^2-W^2)^2$         & \\
          & $m_{xz}=0$, $J_{x \theta}=0$, $J_{z\theta}=0$ $\displaystyle\vphantom{\frac{A}{A}}$&   \\
          \hline
      \cite{Burton2012} $\displaystyle\vphantom{\frac{A}{A}}$ & $m_{xx}=0$                 & $\vec F_{d_B} \cdot \vec e_x=-\nu_x |\dot{x}|^2 \mathrm{sign}(\dot{x})$\\
            & $ m_{zz}=0$                 & $\vec F_{d_B} \cdot \vec e_y=-\nu_y |\dot{y}|^2 \mathrm{sign}(\dot{y})$ \\
           & $I_{\theta \theta}=0$                               & $\vec M_{d_B}\cdot\vec e_z=-\nu_z |\dot{\theta}|^2 \mathrm{sign}(\dot{\theta})$ \\
          & $m_{xz}=0$, $J_{x \theta}=0$, $J_{z\theta}=0$ &   
  \end{tabular}
\end{center}
\caption{Dynamic fluid forces proposed by  \cite{Tsai2007} and \cite{Burton2012}  for iceberg capsize modelling}
\label{tab:previous_models}
\end{table}

\subsection{SAFIM}

To better reproduce the motion of capsizing icebergs, we have developed a new model along the lines of previous propositions by our group \cite{Sergeant2018,Sergeant2019}. The model is referenced as SAFIM for Semi-Analytical Floating Iceberg Model.
In addition to the previously used drag formulation, this new model uses a drag coefficient varying with the aspect ratio and integrates a tunable added-mass effects.
As will be shown, SAFIM reproduces the main results of ISIS-CFD.
For example, it predicts the horizontal movement of capsizing icebergs initially at rest.

A particular feature of the drag model in SAFIM is that it is based on the \emph{local} velocity of the iceberg's surface through which the drag pressure could be approximated as $p = \rho_w v_n \mbox{sign}(v_n)/2$, where 
 $\vec v$ is the local velocity of the iceberg surface, $v_n=|\boldsymbol{v}\cdot \boldsymbol{n}|$ is the corresponding normal velocity, and $\vec n$ is the local normal to the surface of the iceberg. Therefore, the total drag force and the drag torque are evaluated as integrals of local pressure over the submerged part of the iceberg $\Gamma_s$:
 
\begin{equation}
\vec F_d  = -  \displaystyle\alpha \frac12\int\limits_{\Gamma_s} \rho_w v_n^{2}\mbox{sign}(v_n)\vec{n}\,d\Gamma,
\label{eqn:DragSAFIM1}
\end{equation}
\begin{equation}
\vec M_d =- \displaystyle\alpha \frac12\int\limits_{\Gamma_s}\rho_w v_n^{2} \mbox{sign}(v_n) (\vec r - \vec r_G) \wedge \vec{n}\,d\Gamma,
\label{eqn:DragSAFIM2}
\end{equation}
where , $\vec r$ is the local position vector and $\vec r_G$ is the position vector of the iceberg's centre of mass; the wedge sign $\wedge$ denotes a vector cross product.
We consider here a quadratic dependence of the local drag force on the normal velocity, this is discussed in Section~\ref{sec:pressureloc}. The integral expressions for the drag force and torque are given in Appendix~\ref{sec:expressions}.
The factor $\alpha$ of the order of unity is the only adjustable parameter of the drag model. It can be adjusted with respect to the reference CFD simulations or experiments. We recall that in the original papers of~\cite{Sergeant2018,Sergeant2019}, this factor was set to $\alpha=1$ by default. 
However, due to the complexity of the fluid flow, the optimal value of $\alpha$ may change with the geometry of the iceberg.

This formulation, rather than attempting to describe the local pressure accurately, which is difficult based on geometrical considerations only (section~\ref{sec:pressureloc}; \cite{Sergeant2018}), aims at providing a good approximation of the global forces and torques acting on the rotating iceberg.
As opposed to the simplified drag model of~\cite{Burton2012} - in which the drag force and torque depend only on the velocity of the centre of gravity - the hydrodynamic forces $\vec F_{d}$ and the torque $\vec M_{d}$ depend here on the iceberg's current configuration $z_G,\,\theta$ (which determines the submerged part), and on the three velocities $\dot x_G,\,\dot z_G,\,\dot\theta$, (which together with the inclination angle $\theta$ determine the local normal velocity).
This makes it possible to produce a horizontal force acting on the iceberg during capsize even for icebergs initially at rest.
Another advantage is that a unique fit-parameter $\alpha$ is required to represent the drag effect, contrary to three independent fit parameters used in~\cite{Burton2012}.
This makes it possible to easily generalize the model to more complex iceberg geometries.

As for the added-masses in SAFIM, we will consider two possibilities:
\textit{simplified} added-masses and \textit{computed} added-masses.
The simplified added-masses option is based on analytical formulas for the diagonal terms of the added-mass matrix. The coupled terms of added-mass are taken to be equal to zero:  $m_{xz}=0$, $J_{z \theta}=0$, $J_{x \theta}=0$.
The formulas used are taken from~\cite{Wendel1956} for fully or partly submerged solids and were adapted to a capsizing body.
The horizontal and vertical added-masses take the following forms:
\begin{eqnarray}
\label{eqn:AMSAFIM}
m_{xx} &= \displaystyle \frac{1}{4}\,C_{x}\,\pi\,\rho_w\,H_{\mathrm{eff}}^2(z_G,\theta)\\
m_{zz} &= \displaystyle \frac{3}{16}\,C_{z}\,\pi\,\rho_w\,W_{\mathrm{eff}}^2(z_G,\theta)
\end{eqnarray}
where $H_{\mathrm{eff}}$ and $W_{\mathrm{eff}}$ are the effective height and width defined as the projection of the submerged part of the iceberg along the horizontal and vertical axes, respectively (see Fig.~\ref{fig:schema_iceberg_dz0} and Appendix~\ref{sec:expressionsAM}) which depend on the vertical and angular positions of the iceberg. Therefore, the added masses $m_{xx}$ and $m_{zz}$ evolve during the capsize.
On the other hand, the added moment of inertia $I_{\theta \theta}$ is assumed to depend only on the height of the iceberg, so it remains constant during the capsize:
\begin{eqnarray}
\label{eqn:AISAFIM}
I_{\theta \theta} &=\displaystyle 0.1335\,C_{\theta}\,\pi\,\rho_w\,\left( \frac{H}{2} \right) ^4 .
\end{eqnarray}
In order to adjust the added-mass effect used in SAFIM to reproduce the reference ISIS-CFD results, we introduce three calibration factors in the above equations: $C_{x}$, $C_{z}$ and $C_{\theta}$ (see Section \ref{sec:Results} for calibration).

\textit{Computed} added-masses are calculated using a computational fluid dynamics solver. This is done by applying a unit acceleration on the iceberg for the considered degree of freedom, which leads to a simplified expression of the Navier-Stokes equations (eq. (16) in \cite{Yvin2018}). Then we obtain an equation for the pressure (the eq. (18) in \cite{Yvin2018}) which can be solved on the fluid domain using a numerical method such as the finite element, boundary element method or finite-volume method (used in this study \cite{Queutey2007}).
The integration of the induced pressure on the emerged part of the iceberg in response to a unit acceleration along $x$, $z$ or rotation around $y$ gives the corresponding column of the symmetrical added-mass matrix including both diagonal and coupled entries.
Similarly to the simplified added-mass, the values of the computed added-mass also depend on the iceberg position and they therefore evolve during the capsize. 
For the computed added-masses, the coupled terms are non-zero, giving rise to a coupling between horizontal, vertical and rotational accelerations.

To solve the motion equations~(\ref{eqn:alleq1}-\ref{eqn:alleq3}) with SAFIM, the St\"ormer-Verlet integration scheme is used.
Since SAFIM has only three degrees of freedom, the integration over time is very fast, only a few seconds compared to few hours for ISIS-CFD on a single CPU. 
The time step in SAFIM that ensures a sufficiently accurate results is $\Delta t = 0.1$~s in the field scale and $\Delta t = 0.001$~s in the laboratory scale. 
In both cases, this {step corresponds to a dimensionless time step of $\Delta t' =\Delta t/ \sqrt{H/g} \approx 0.01$.

\section{Performance and limitations of SAFIM}
\label{sec:Results}

\subsection{SAFIM's calibrations}
\label{sec:Studiedcasesandoutputs}

The validation of the proposed model should be suited to the final objectives:
(1) an accurate reproduction of the forces exerted by the water on the iceberg during capsize;
(2) the ease of implementation in a finite element solver for simulation of the whole iceberg-glacier-bedrock-ocean system (left for future work),
(3) suitability of the model for the entire range of possible geometries of icebergs encountered in the field.
In this context, we consider 2D icebergs with rectangular cross-sections (Fig.~\ref{fig:schema_iceberg_dz0}). We use typical densities observed in the field. As discussed in Section~\ref{sec:disc}, the considered density has a non-negligible effect on the calving dynamics.
We apply SAFIM to the same four geometries as described in Section~\ref{sec:Burton}, with initial tilt angles given in Section~\ref{sec:complab} and Table~\ref{tab:err}.

To compare SAFIM and ISIS-CFD results, we compute the mismatch in the time-evolution of the horizontal forces $F_x$ ($\mathcal{L}^2$ norm) during the capsize. The phases of the capsize that we focus on are phases (ii) and (iii) (defined in Section~\ref{sec:lab_scale}). The reason we do not seek to perfectly model the initial phase (i) with SAFIM is discussed in Section~\ref{sec:time}. Also, SAFIM is designed to model the capsize phase but cannot model the very end of the capsize ($\theta>80\degree$), i.e. phase (iv), nor the post-capsize phase (v). In these phases, forces induced by complex fluid motion, which are difficult to parametrize, are expected to dominate gravity and buoyancy forces.

For SAFIM with a drag force and no added-masses, the mismatch is defined as:
\begin{equation}
\displaystyle E_1 = \frac{ \int_{t_{1}}^{t_{2}} \left|F_{x_{\text{ISIS}}}(t) - F_{x_{\text{SAFIM}}}(t-\Delta t)\right|^2 \text{d}t} {\int_{t_{1}}^{t_{2}} |F_{x_{\text{ISIS}}}(t)|^2 \text{d}t}
\label{eq:erreuri}
\end{equation}
with $F_x$ being the total horizontal force acting on the iceberg, $t_{1}$ such that $F_x(t_{1})=1/6 F_{\min}$ and $t_{1}<t_{\min}$ and $t_{2}$ such that $F_x(t_2)=1/6 F_{\min}$ and $t_{2}>t_{\min}$ , with $F_{\min}$ being the first extremum of the force and $t_{\min}$ the time at which it occurs. In Fig.\ref{fig:Fx_T1_L1_ISIS}, $t_1'=10.3$ and $t_2'=14.2$. This time interval is within phases (ii) and (iii).
Since without added-mass the initial phase of capsize cannot be accurately reproduced, for the comparison purpose, the SAFIM response is shifted artificially in time so that the first extremum of the curve fits that of ISIS-CFD. This time shift $\Delta t$ is discussed in Section~\ref{sec:time}.

For the SAFIM model with a drag force and \emph{added-masses}, a more demanding mismatch is used, which reads as:
\begin{equation}
E_2 = \frac{ \int_{0}^{t_{3}} \left|F_{x_{\text{ISIS}}}(t) - F_{x_{\text{SAFIM}}}(t)\right|^2 \text{d}t} { \int_{0}^{t_{3}} |F_{x_{\text{ISIS}}}(t)|^2 \text{d}t}
\label{eq:erreurii}
\end{equation}
where $t=0$ is the beginning of the simulation and $t_{3}$ the first time for which the horizontal force crosses zero $F_x(t_{3})=0$ after $t_{\min}$, which corresponds to $t_3' = 14.4$ in Fig.\ref{fig:Fx_T1_L1_ISIS}. This time interval includes phases (i), (ii) and (iii). The force is not shifted in time as for $E_1$.
Errors~(\ref{eq:erreuri}) and (\ref{eq:erreurii}) are computed for a parametric space of drag and added-mass coefficients as detailed in Appendix~\ref{sec:tabparam}. 
The optimized values of the coefficients $\alpha$, $C_{x}$, $C_{z}$ and $C_{\theta}$ are chosen such that the errors $E_1$ and $E_2$ are minimized.

\subsection{Effect of drag and added-mass}
\label{sec:anapara}
We analyse the horizontal force produced by SAFIM in order to understand the effects of the drag force and added-masses on the dynamics of the iceberg capsize.
To do so, the dynamics of an iceberg with an aspect ratio $\varepsilon=0.246$ were simulated by SAFIM with and without drag and with and without added-masses and compared to the results of ISIS-CFD (Fig.~\ref{fig:courbes}) :

\begin{enumerate}
\item[\textit{case 0:}] no drag and no coupled terms in the added-mass matrix: the horizontal force predicted by SAFIM is equal to zero $F'_x=0$, $\forall~t$, as expected. For the sake of clarity, this case is not plotted in Fig.\ref{fig:courbes}.

\item[\textit{case 1:}] drag and no added-mass (no AM) is shown in Fig. 7 for two values of the drag coefficient $\alpha=1$ (purple curve) and $\alpha=0.85$ (blue curve). The value $\alpha=0.85$ is the optimized value of $\alpha$ obtained by minimizing the error $E_1$.
The force has a slightly higher amplitude (around $t'\approx 8.5$) and duration with $\alpha=1$ than with the optimized drag coefficient $\alpha=0.85$. Even though the amplitude and shape of the SAFIM horizontal force are very similar to the ISIS-CFD results, the full capsize occurs earlier with SAFIM.
When the SAFIM curves are shifted in time by $\Delta t'=2.7$ (cyan curve in Fig.~\ref{fig:courbes}), the previous SAFIM force fits well ISIS-CFD. The shape is the same and the error on the waveform is $E_1=5.2\%$, with a relative error on the first force extremum of $4\%$. A comparison of SAFIM with the optimized $\alpha$-factor, SAFIM with $\alpha=1$, and ISIS-CFD is given in Appendix \ref{sec:alpha1}, for the four aspect ratios.

\item[\textit{case 2:}]
drag and simplified added-mass: when coefficients for the drag and added-masses are taken all equal to $1$ the horizontal force is very different from the reference ISIS-CFD results (orange curve in Fig.~\ref{fig:courbes}). The duration of the capsize is largely over-estimated and the amplitude is strongly underestimated.
The optimized drag and added-mass coefficients that give a minimum error $E_2$ are $\alpha=1.1$ for the drag, $C_{\theta}=0.75$ for the added moment of inertia and zero factors $(C_x,C_z)=(0,0)$. 
The corresponding results (yellow curve in Fig.~\ref{fig:courbes}) are in a very good agreement with the reference results ($E_2=10\%$) both for the shape and for the time corresponding to the force extremum $t' \approx 11.45$. The added moment of inertia (coefficient $C_{\theta}$) slows down the initial rotation of the iceberg. However, the amplitude of the force extremum is slightly underestimated by $8\%$. The accuracy of the formula of the simplified added-masses with coefficients equal to $1$ is discussed in section~\ref{sec:AM}.

\item[\textit{case 3:}] 
no drag and computed added-masses: SAFIM (dark green curve in Fig.~\ref{fig:courbes}) fits the reference results quite well in amplitude but not in time and predicts a huge second minimum.

\item[\textit{case 4:}]
drag and computed added-masses: when correcting the drag coefficient to $\alpha=3.0$, which minimizes the error $E_2$, SAFIM fits better in time, reproducing the initial slow change of the force, but the amplitude and the shape still do not fit ISIS-CFD (black curve in Fig.~\ref{fig:courbes}). Similarly to the simplified added-masses, the computed added-masses slow down the initial rotation of the iceberg.
\end{enumerate}

\begin{figure}[htb!]
\centering
\includegraphics[width=1\linewidth]{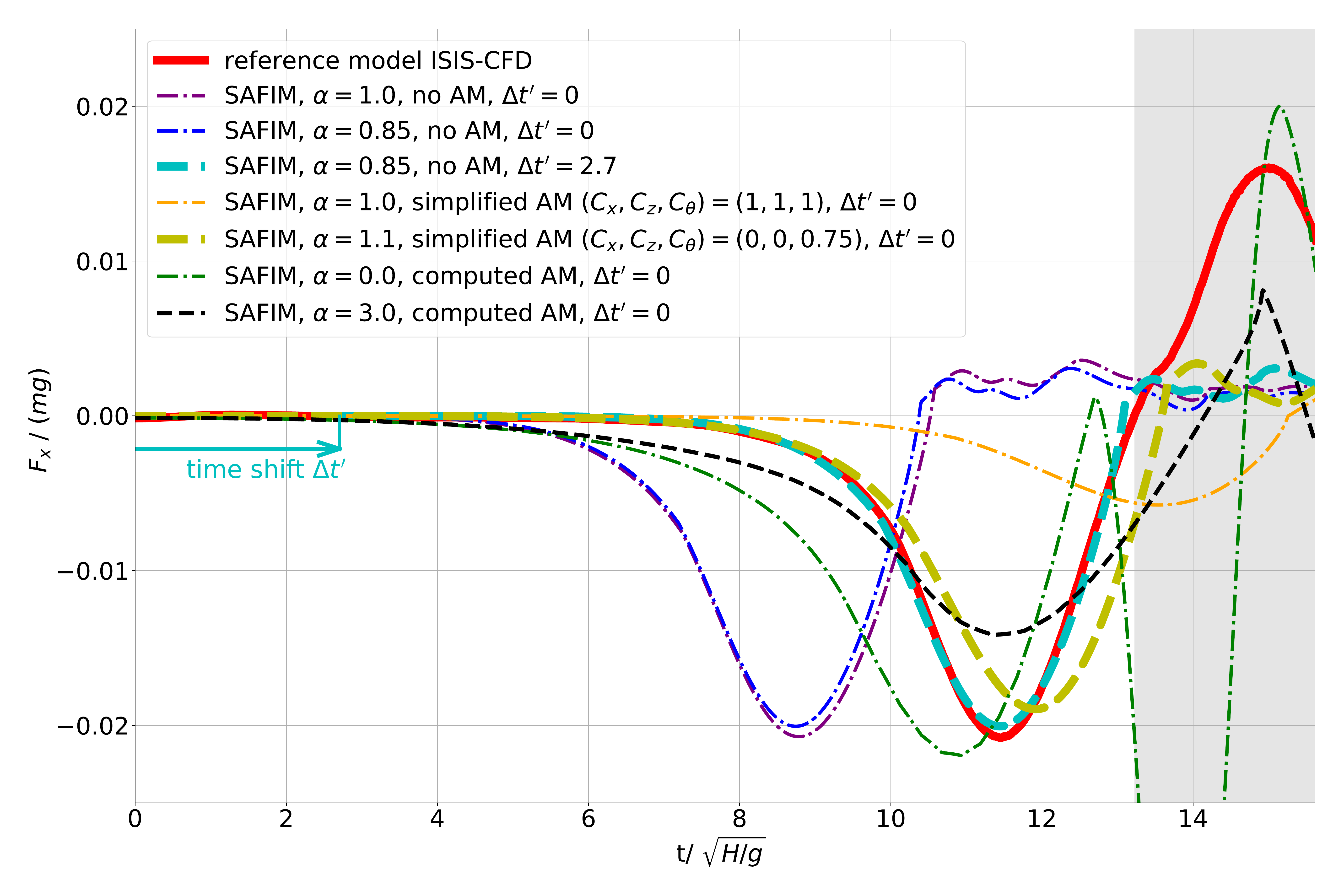}
\caption{Computed horizontal force $F'_x=F_x/(mg)$ applied on the iceberg versus dimensionless time $t'=t/ \sqrt{H/g}$ for a reference ISIS-CFD simulation and for SAFIM simulations with different drag and added-mass parameters.}
\label{fig:courbes}
\end{figure}

This analysis suggests that the drag force has mainly an effect on the amplitude of the first force extremum and that the added-masses have an effect on the duration of the initiation of the capsize. Also, optimized coefficients of drag and added-masses improve the model significantly compared with the case with all coefficients set to $1$. Further discussions on the pros and cons of the SAFIM models are given in Section~\ref{sec:dispref}.

\subsection{Effect of the iceberg's aspect ratio}
\label{SAFIM}

We will now analyse the forces and the torque acting on the four selected geometries of icebergs computed by ISIS-CFD and SAFIM.
The evolution of the dimensionless horizontal force $F'_x$, vertical force $F'_z$, torque $M'_{\theta}$, horizontal displacement $x'_G$, vertical displacement $z'_G$ and inclination $\theta$ obtained by ISIS-CFD and SAFIM are plotted in Fig.~\ref{fig:allcurvesnoAM} for SAFIM best-fitted results obtained with drag and without added-masses and in Fig.~\ref{fig:allcurvesAMSAFIM} for SAFIM results with drag and simplified added-masses. SAFIM models use optimized parameters indicated in Table~\ref{tab:err} for each aspect ratio.

We will first discuss the sensitivity of the forces computed with the reference model ISIS-CFD to the aspect ratios $\varepsilon$. We observe that the amplitude of the first extremum of both the horizontal force $F'_x$ and the vertical force $F'_z$ decreases with increasing aspect ratio.
Consequently, the amplitude of the horizontal acceleration $\ddot x_G=F_x / m= g F'_x$, also decreases with increasing aspect ratio. This is consistent with the observed slower horizontal displacement of icebergs with larger aspect ratios as reported in Section~\ref{sec:complab}.
Also, the durations of the capsize do not differ much in the four cases. 
\begin{itemize}
\item \textit{drag and no added-masses (case 1):} The minimal error $E_1$ increases with the aspect ratio (from $5\%$ for $\varepsilon=0.246$ and up to $24\%$ for $\varepsilon=0.639$).
The optimal drag coefficient $\alpha$ increases in an approximately affine way with the aspect ratio (Fig.~\ref{fig:err}a) as:
$$\alpha_{\text{opt}}(\varepsilon)\approx-1.6+8.8 \varepsilon$$
with the coefficient of determination equal to $R^2=0.98$.
This linear regression is valid within the range of studied aspect ratios $0.246\le \varepsilon\le 0.639$. Note that this formula should not be used for $\varepsilon < 0.18$ for which the drag coefficient would be negative, which is physically meaningless.
\item \textit{drag and simplified added-masses (case 2):} The minimal errors $E_2$ for SAFIM with drag and simplified added-masses are greater than the errors $E_1$ with drag and no added-masses Fig.~\ref{fig:err}(b).
As for $E_1$, the error $E_2$ increases with the aspect ratio (from $10\%$ for $\varepsilon=0.246$ up to $26\%$ for $\varepsilon=0.639$).
Note that for all the four studied cases, optimization of the error requires keeping only one non-zero added-mass coefficient, namely the added moment of inertia coefficient.
The simplified added-masses allows a slow initiation of rotation, which can be explained by an added moment of inertia of the surrounding fluid.
\item \textit{drag and computed added-masses (case 4):} In that case, SAFIM predicts the time and amplitude of the extremum of the force and the torque less accurately than the two previous cases: the error $E_2>34\%$ for all the four studied cases (Fig.~\ref{fig:err}b). The corresponding results are not shown here.
\end{itemize}
\begin{figure}[htb!]
\centering
\includegraphics[width=1\linewidth]{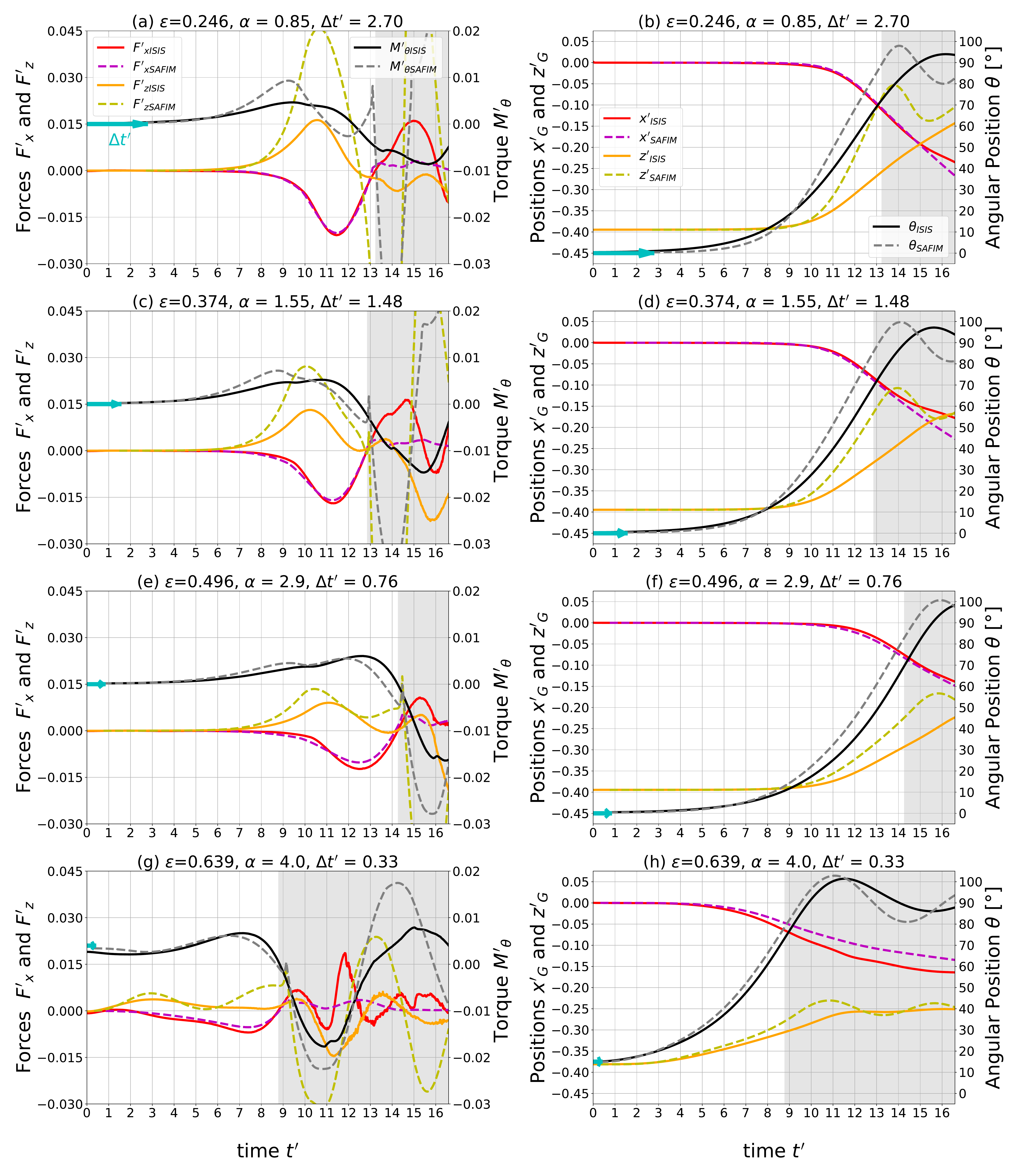}
\caption{Capsize simulations for SAFIM with drag, without added-masses and with time shits, and for ISIS-CFD: evolution of the dimensionless total horizontal force $F'_x$, vertical force $F'_z$ and torque $M'_{\theta}$ on the iceberg (a, c, d, e), of the horizontal $x'_G$ and vertical $z'_G$ positions of $G$ and of the inclination $\theta$ of the iceberg (b, d, f, g). Results are given for icebergs with $\varepsilon=0.246$ (a, b), $\varepsilon=0.374$ (c, d), $\varepsilon=0.496$ (e, f) and $\varepsilon=0.639$ (g, h). SAFIM curves were shifted (green arrow) by the dimensionless time $\Delta t'= \Delta t \, \sqrt{g/H}$. The SAFIM drag coefficient $\alpha$ and time $\Delta t'$ are indicated in the titles.}
\label{fig:allcurvesnoAM}
\end{figure}

\begin{figure}[htb!]
\centering
\includegraphics[width=1\linewidth]{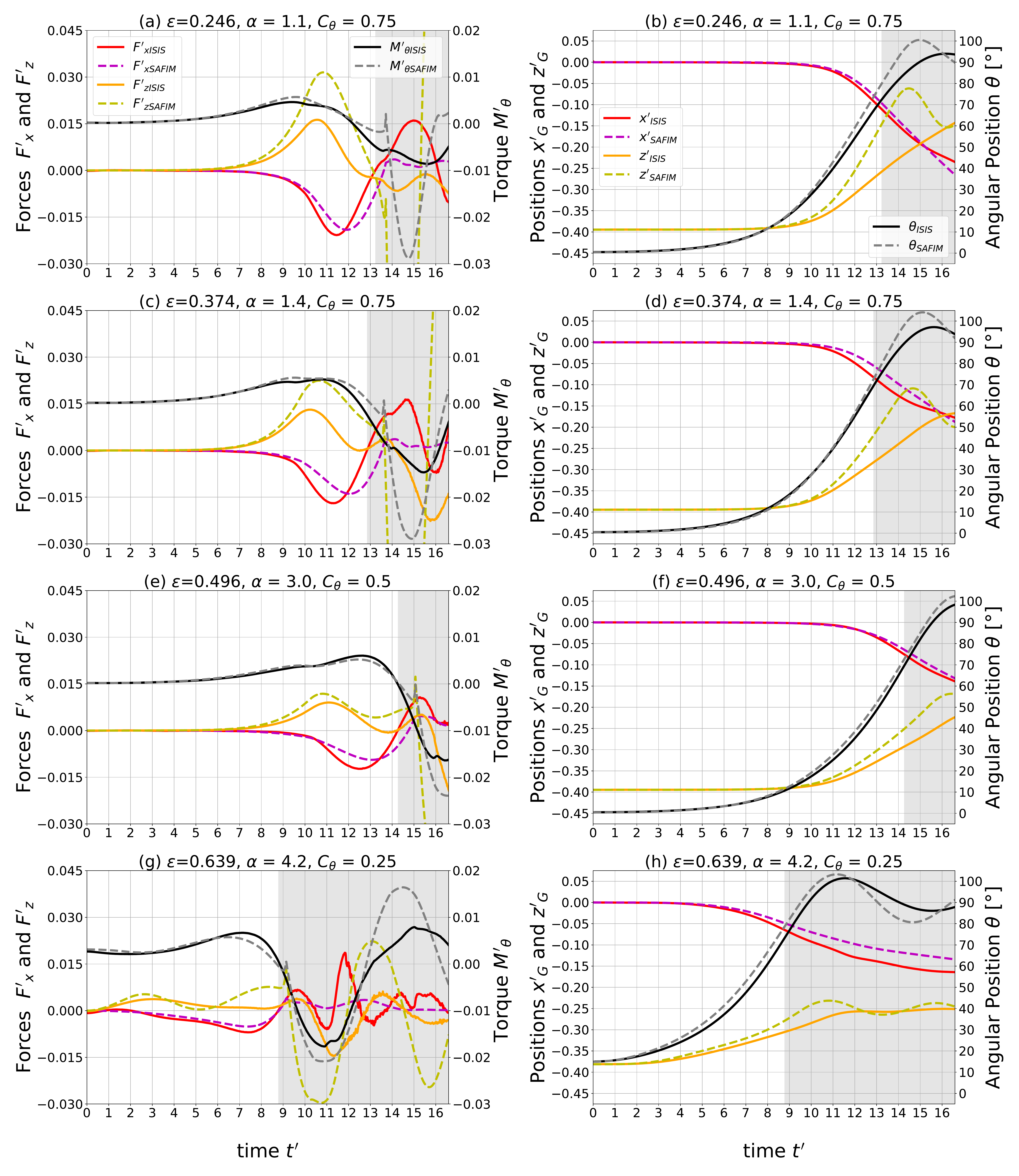}
\caption{Same as in Fig.~\ref{fig:allcurvesnoAM} but for SAFIM with drag, simplified added-masses and no time shift ($\Delta t'$). SAFIM drag coefficient $\alpha$ and added-mass coefficient $C_{\theta}$ are indicated in the titles.}
\label{fig:allcurvesAMSAFIM}
\end{figure}

\begin{figure}[htb!]
\centering
\includegraphics[width=1\linewidth]{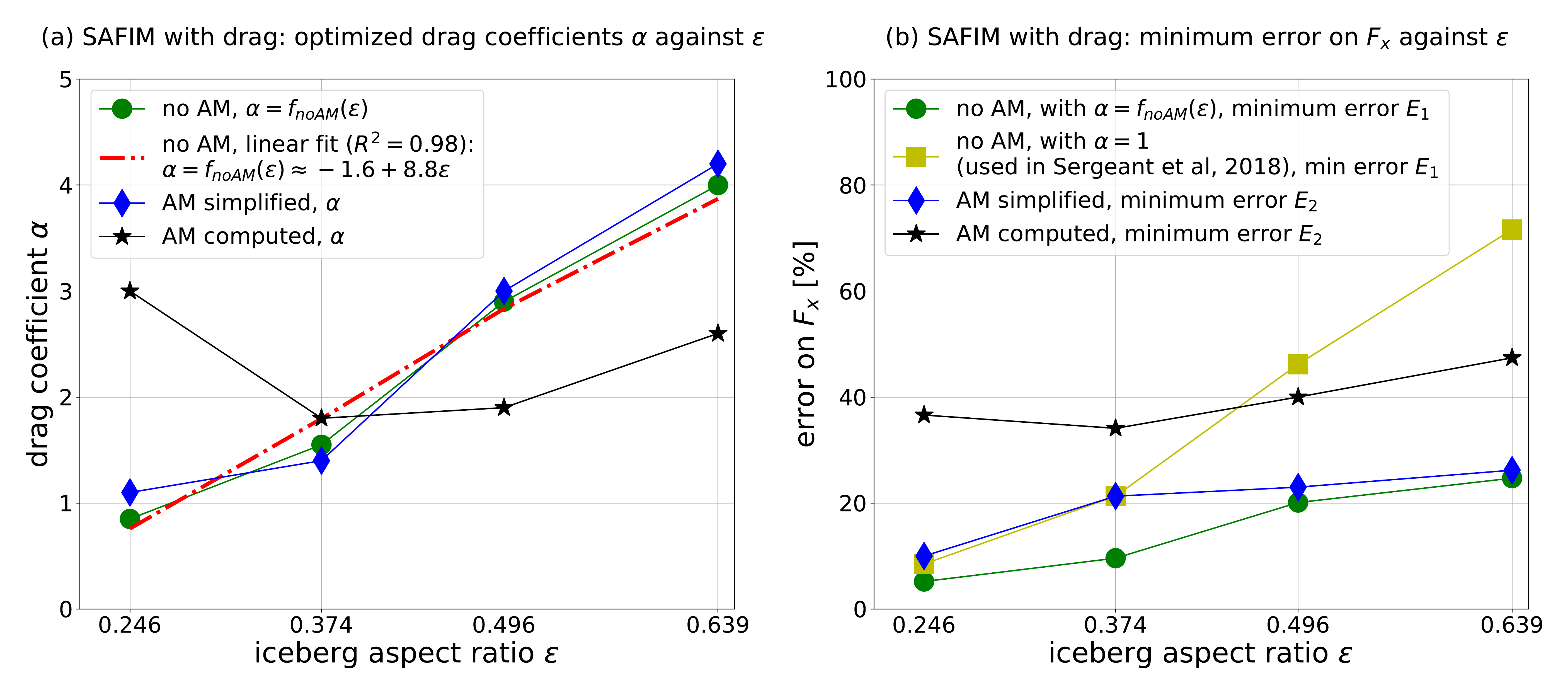}
\caption{(a) Optimized values of the drag coefficient $\alpha$ for different iceberg's aspect ratios, which were determined for SAFIM with and without added-mass, only added moment of inertia was used in the simplified added-mass model, and a full added-mass matrix was used for the computed added-mass;
(b) the minimal error of the horizontal force corresponding to different models for different iceberg's aspect ratios. The optimal parameters and corresponding minimal errors are also given in Table~\ref{tab:err}}
\label{fig:err}
\end{figure}


\section{Discussion}
\label{sec:disc}

First we will discuss the performance of SAFIM in Sections~\ref{sec:dispref} and \ref{sec:time}, then the modelling choices in Sections~\ref{sec:pressureloc} and \ref{sec:AM} and finally the sensitivity of the model results to geophysically meaningful variations of parameters in Sections~\ref{sec:density}, \ref{sec:3d} and \ref{sec:disgeom}.

\subsection{SAFIM performance and comparison with existing models\label{sec:dispref}}

The advantages of the formulated and validated SAFIM model with drag and without added-masses is that (1) it can be readily implemented in a finite element model like the one in \cite{Sergeant2018}, (2) it requires only one parameter, the drag coefficient $\alpha_{\text{opt}}(\varepsilon)\approx-1.6+8.8 \varepsilon$,
(3) it quite accurately reproduces the shape and amplitude of the horizontal force.
The drawback of this model is that it does not correctly simulate the kinematics of the iceberg capsize, especially the time needed to reach the peak force (see discussion in Section~\ref{sec:time}). In addition, the evolution of the torque and vertical force is not well reproduced.

The advantage of SAFIM with drag and simplified added-masses is that it correctly reproduces the time of the force extremum (no shift in time is needed) and it reproduces the torque and the vertical force better than SAFIM with drag and no added-masses. Its drawback is that it underestimates the amplitude of the first extremum of the horizontal force by $\approx10\%$.

SAFIM with drag and computed added-masses gives less accurate results than the two other versions. Assuming that the computed added-masses are physical and accurate~\cite{Yvin2018},
the drag model in SAFIM is not suitable with that added-mass formulation since it does not make it possible to reproduce the dynamics of the iceberg.

The proposed SAFIM model well predicts the first part of the horizontal force applied by the fluid on the iceberg, either when using a drag force only (i.e. no added-masses) and shifting the curve in time or when using a drag force and simplified added-masses (and no shift in time). However, the evolution of the force after the capsize ($\theta>80\degree$) is not well modelled. This is probably due to the fact that the evolution of the local fluid pressures is governed by a complex fluid motion around the iceberg (see Section~\ref{sec:pressureloc}) which is hard to parametrize without full fluid dynamics computations. The duration and amplitude of the positive peak in the force is however comparable to that of the first minimum of the force (see e. g. for $\epsilon=0.374$ Fig.\ref{fig:allcurvesnoAM}).

An advantage of SAFIM over previous models~\cite{Tsai2008,Burton2012} is that, thanks to a special form of the drag force, it can describe the horizontal movement of a capsizing iceberg triggered by its rotation. As shown in~\cite{Sergeant2018}, SAFIM can distinguish between a top-out and bottom-out capsize, when used to simulate the contact force between a capsizing iceberg and a rigid glacier front. A qualitative explanation of the emerging non-zero horizontal force given by the drag force is given in Appendix~\ref{sec:expressions}.

We calculate the error in SAFIM with a drag coefficient $\alpha=1$ for all aspect ratios and without added-mass (Fig.~\ref{fig:err}). Exactly the same model was used in \cite{Sergeant2018, Sergeant2019}, but for modelling an iceberg capsizing in contact with a glacier front. The error $E_1$ is about two times greater than when taking the optimum value of the $\alpha$ coefficient for each aspect ratio, and the amplitude and duration of the first negative part of the force is underestimated - except for the thinnest iceberg with $\varepsilon=0.246$, for which the opposite is true. Note that this error is only relevant for a freely capsizing iceberg. In future work, the errors for an iceberg capsizing in contact with a glacier front should be estimated but this will require a reference model for fluid-structure interactions that can track the contact between solids, which is a challenging problem~\cite{mayer20103d}.

\subsection{Initiation of the capsize}
\label{sec:time}

In the previous sections, the drag parameter for SAFIM without added-masses was optimised by implementing an artificial time shift of the SAFIM force curve with respect to ISIS-CFD.
This was done because, as already mentioned, SAFIM without added-masses is not able to predict the accurate duration of the initiation of the capsize, where the motion is slow and the horizontal force is close to zero.

Various reasons suggest that this initial phase may not be relevant in the global context of the ultimate objective of the project, i.e. estimation of the short time-scale volume loss on marine-terminating icebergs. To achieve this objective, we need to compare the modelled contact force with the inverted seismic source force. The very beginning of the seismic force has a too small signal-to-noise ratio, therefore it is the first peak of the force that is used as a reference to compare the seismic force and the modelled force. Also, because this force evolves very slowly at the beginning, it will not be responsible for the generation of seismic waves with a period of $50$ s that is predominantly observed on glacial earthquake seismograms \cite{Ekstrom2003, Tsai2007, Tsai2008, Sergeant2018}. Another reason for ignoring the beginning of the capsize is that the duration of the initial slow rotation (phase (i)) of the iceberg is strongly dependent on the initial angle of inclination of the iceberg which is hard to constrain in the field data and, when it is sufficiently small, has little effect on the capsize (phases (ii), (iii), (iv)). The initiation phase of the capsize may also depend on the asymmetrical geometry of the iceberg, its surface roughness and the 3-D effects (see sections \ref{sec:3d} and \ref{sec:disgeom}).

Nevertheless, if we consider a complete  glacier / ocean / bedrock / iceberg / ice-m\'{e}lange system, the initial detachment of the iceberg can result in various other effects such as basal sliding or vertical oscillations of the glacier tongue, which can produce a seismic signal. Therefore the superposition of these phenomena can be erroneous if the timing is not well reproduced.
To solve this issue, simulations of the complete glacier / ocean / bedrock / iceberg / ice-m\'{e}lange system with a full fluid dynamics model coupled with a model for dynamics of deformable solids would seem to be unavoidable, however, as already discussed, it lies beyond actual computational possibilities of the softwares that we dispose.

\subsection{Drag force and local pressure field}
\label{sec:pressureloc}
Following~\cite{Burton2012}, a linear drag model with a local pressure proportional to the normal velocity $|v_n|$ was also tested in SAFIM.
It results in the following modification to equations~(\ref{eqn:DragSAFIM1}) and~(\ref{eqn:DragSAFIM2}):
\begin{equation}
 \vec F_d  = -  \displaystyle\alpha \frac12\int\limits_{\Gamma_s} \rho_w |v_n|\mbox{sign}(v_n)\vec{n}\,d\Gamma,
\end{equation}
\begin{equation}\vec M_d =- \displaystyle\alpha \frac12\int\limits_{\Gamma_s}\rho_w|v_n| \mbox{sign}(v_n) (\vec r - \vec r_G) \wedge \vec{n}\,d\Gamma.
\end{equation}
Such a drag model yields worse results than the original model with quadratic dependency when compared with the reference ISIS-CFD model.
In addition, other drag models were tested with linear and quadratic pressure dependency on the velocity, with a non-uniform parameter $\alpha$ on the surface of the iceberg and with drag depending on the sign of the local normal velocity $v_n$.
Of all drag models tested, the most accurate was the model with quadratic dependency on the normal velocity and with a constant $\alpha$-factor over the whole surface of the iceberg.
However, to better fit the reference results, the $\alpha$-factor was made dependent on the iceberg's aspect ratio, which is an important difference with the original model presented in~\cite{Sergeant2018}.

To go further in our understanding of the forces generated by the fluid, we analyse the hydrodynamic pressure distribution on the sides of the iceberg, computed by ISIS-CFD and defined as $p_{dyn}=p_{tot}-p_{sta}$,
with $p_{tot}$ the total fluid pressure and $p_{sta}$ the hydrostatic pressure computed for the reference still water level ($z=0$).
In particular, we attempted to establish a link between the spatial distribution of the hydrodynamic pressure on the iceberg and the local features of the fluid flow, notably with the normalized vorticity (see Fig.~\ref{fig:Fx_T1_L1_ISIS}) which is defined as: $\omega = -\sqrt{H/g}\;\vec e_y \,\cdot (\nabla \wedge \vec u),$ with a negative value (blue) accounting for a vortex rotating clockwise and a positive (red) value for a counter-clockwise vortex.
On the four snapshots presented in Fig.~\ref{fig:Fx_T1_L1_ISIS}, we also plot the dimensionless hydrodynamic pressure $p'_{dyn}=p_{dyn}/ (\rho_i\,H\,g)$. The hydrodynamic pressure is plotted as a shaded pink area outside the iceberg for a negative pressure and inside the iceberg for a positive pressure. Note that these values are about two orders of magnitude lower than the average hydrostatic pressure.
The dynamic pressure is higher at locations where there is a vortex close to the surface of the iceberg such as on the corner furthest right in Figs.~\ref{fig:Fx_T1_L1_ISIS}(b) and \ref{fig:Fx_T1_L1_ISIS}(c), on the bottom part of the left side and in the middle on the right side of the iceberg.
This observation suggests that the dynamic pressure field is highly dependent on the vortices in the fluid. Such an evolution of complex vortex motion cannot be reproduced within SAFIM and requires the resolution of the equations of fluid motion as in ISIS-CFD. Note that the high values of the pressure on the top side of the iceberg in Figs~\ref{fig:Fx_T1_L1_ISIS}(c) and \ref{fig:Fx_T1_L1_ISIS}(d) are due to an additional hydrostatic pressure produced by the wave that is above the reference sea level.

Using ISIS-CFD simulations, we made an attempt to correlate the local hydrodynamic pressure $p_{dyn}$ with the normal velocity $v_n$ via a power law as is the case in SAFIM:
\begin{equation}
|p_{dyn}|=b|v_n|^a.
\label{eq:pa}
\end{equation}
in order to optimize the drag law used here (in SAFIM, the coefficients are $a=2$ and $b=-\alpha\,\mathrm{sign}(v_n)\,\rho_w/2$).
This attempt was not successful. We observed that the values of $a$ and $b$ vary significantly along the sides of the iceberg and with time particularly on the top part of the long sides of the iceberg and close to the corners.
Also, we tried to correlate the dynamic pressure for $a=2$, as in SAFIM, without success. Nevertheless, the choices made in SAFIM ensures rather accurate overall drag forces and torques acting on the iceberg due to dynamic pressure.

\subsection{Accuracy of the added-mass}
\label{sec:AM}

The simplified added-masses, defined by Eqs. (\ref{eqn:AMSAFIM}), (8), (\ref{eqn:AISAFIM}), with only diagonal terms in the added-mass matrix, will now be compared with the reference computed added-masses.
Both added-mass matrices depend on the current iceberg position, and therefore they should be updated at every time step. These matrices are calculated for the iceberg's motion computed by ISIS-CFD.
We show the time evolution of the added-masses and the added moment of inertia, for the capsize of a laboratory-scale iceberg with $H=0.103\,\meter$ and $\varepsilon=0.246$ in Fig.~\ref{fig:am}(a-c) and $\varepsilon=0.496$ in Fig.~\ref{fig:am}(d-f).

\begin{figure}[htb!]
\centering
\includegraphics[width=1\linewidth]{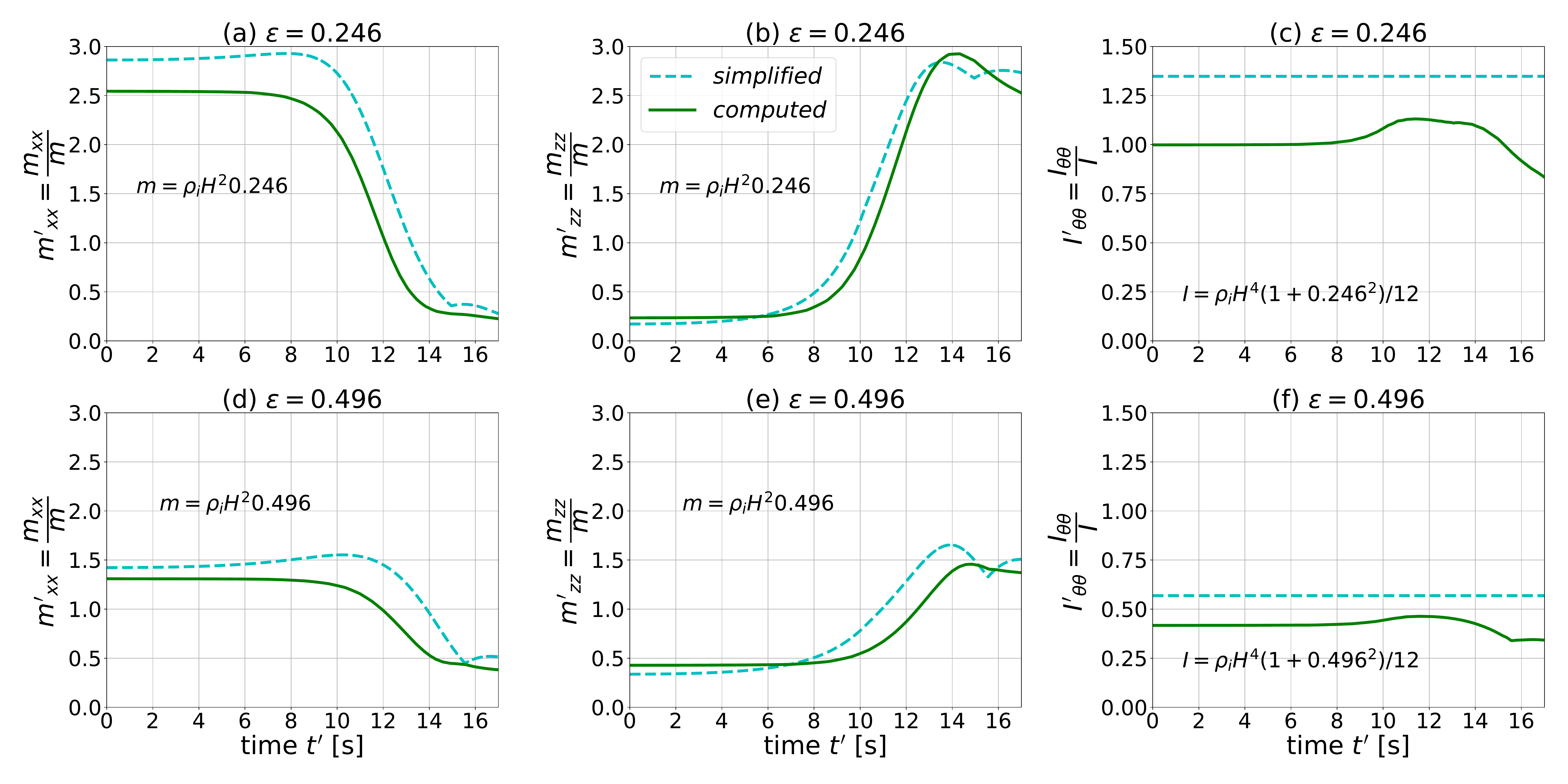}
\caption{Time evolution of dimensionless horizontal added-masses (a) and (d), horizontal added-masses (b) and (e), added moment of inertia (c) and (f). The dashed cyan curves are the simplified added-masses and moment of inertia and the solid green curves are the computed added-masses and moment of inertia. Values are given for a laboratory-scale iceberg ($H=0.103\;\meter$) with field densities and aspect ratio $\varepsilon=0.246$ in (a), (b) and (c) and $\varepsilon=0.496$ in (d), (e) and (f). For each aspect ratio $\varepsilon$, the values are normalized by the mass of the iceberg $m=\rho_i H^2\varepsilon$ and the inertia of the iceberg $I=\rho_i H^4(1+\varepsilon^2)/12$. The non-constant added-masses are given for the positions of the iceberg in the ISIS-CFD simulations. The values of the simplified added-masses are plotted for all coefficients equal to $1$: $C_x=C_z=C_{\theta}=1$.}
\label{fig:am}
\end{figure}
The simplified horizontal and vertical added-masses are in very good agreement with the corresponding computed added-masses: relative error (with the $\mathcal{L}^2$ norm) of $21\%$ on $m_{xx}$ for $\varepsilon=0.246$ and $23\%$ for $\varepsilon=0.496$; relative error of $11\%$ on $m_{zz}$ for $\varepsilon=0.246$ and $13\%$ for $\varepsilon=0.496$.
The simplified added moment of inertia $I_{\theta\theta}$ is assumed to be constant in our model whereas the computed one varies in time and has a smaller value: relative error of about $30\%$.

For the aspect ratio $\varepsilon=0.246$, the horizontal added-mass $m_{xx}$ decreases from $\approx2.5\,m$ at the beginning down to $\approx0.2\,m$, where $m$ is the iceberg mass, whereas, symmetrically, the vertical added-mass $m_{zz}$ increases from $\approx0.2\,m$ at the beginning up to $\approx2.5\,m$ at the end of the capsize.
The horizontal added-mass $m_{xx}$ measures the resistance of the fluid to a horizontal acceleration $\ddot{x_G}$ of the iceberg.
The iceberg has a longer submerged vertical extension (of the order of $H$) at the beginning than at the end (of the order of $W$) of the capsize, thus it needs to displace a greater volume of the fluid in a horizontal motion at the beginning than at the end of the capsize ($\theta>90 \degree$). Therefore, a greater added-mass $m_{xx}$ is expected at the beginning of the capsize.
The vertical added-masses $m_{zz}$, sensitive to the horizontal extension of the iceberg, experience the opposite variations in time.
The computed added moment of inertia is equal to the moment of inertia of the iceberg $I$ at the beginning. Then it increases to $1.25\,I$ and decreases below the iceberg's moment of inertia to $0.9\,I$ (see Fig.~\ref{fig:am}).

For the aspect ratio $\varepsilon=0.496$, the variations of the dimensionless added-masses are different in amplitude but with rather similar evolutions.

For other geometries, the added-masses (resp. inertia) are also of the same order of magnitude as the masses (resp. inertia) of the iceberg as found here for rectangular icebergs. For example, in the case of a two-dimensional thin ellipse with an aspect ratio of $b/a$, with $a$ the along-flow dimension and $b$ the cross-flow dimension, \cite{Newman1999} gives the added-masses and added moment of inertia. For $b/a=0.2$, the transverse added-mass is equal to $0.9$ times the mass of the \textit{displaced volume of fluid} (\textit{i.e.} the submerged volume of the solid times the density of the fluid) and the added moment of inertia is equal to $0.7$ times the inertia of the displaced volume of fluid. For similar densities for the fluid and the solid, the added-masses and the added moment of inertia are close to those of the solid.
The optimized values of the coefficients for the simplified added-masses and moment of inertia in SAFIM are given in Table~\ref{tab:err}. 
These values are not in agreement with the reference computed added-masses. However, as discussed in section~\ref{sec:dispref}, the SAFIM model with the simplified added-mass matrix gives better results than SAFIM with the computed added-mass matrix. For $\varepsilon= 0.246$, note that the optimized simplified added moment of inertia ($C_{\theta} = 0.75$) is close to the computed one. However, the simplified added moment of inertia is not in agreement with the computed one for higher $\varepsilon$.

The optimized value $C_z=0$ is consistent with the choice of $m_{zz}=0$ in \cite{Tsai2008} even though it is not equal to the computed vertical added-mass. The optimized coefficient $C_x=0$ gives $m_{xx}=0$. The horizontal added-mass $m_{xx}$ from \cite{Tsai2008} varies similarly to the computed added-mass. The added moment of inertia $I_{\theta\theta}$ with the formula in \cite{Tsai2008} is constant throughout the capsize and different from the optimized added moment of inertia. However, the formula for added-masses and added moment of inertia from \cite{Tsai2008} were given for the simulation of an iceberg capsizing in contact with a wall, which may significantly affect the values of the added-masses.

\subsection{Effect of water/ice densities}
\label{sec:density}
The laboratory experiments discussed in Section.~\ref{sec:Burton} were conducted with water and ice densities slightly different from those in the field (see Section.~\ref{sec:geom}).

As shown in Fig.~\ref{fig:dens}, the dynamics of the iceberg computed by ISIS-CFD with field densities is significantly different from those obtained with laboratory densities: the amplitude, duration and initiation of the capsize are very sensitive to changes in densities.
This sensitivity is also very well reproduced by SAFIM with drag and no added-masses.
Note that no change in the drag coefficient $\alpha$ is needed to accurately reproduce this effect with SAFIM.

\begin{figure}[htb!]
\centering
\includegraphics[width=1\linewidth]{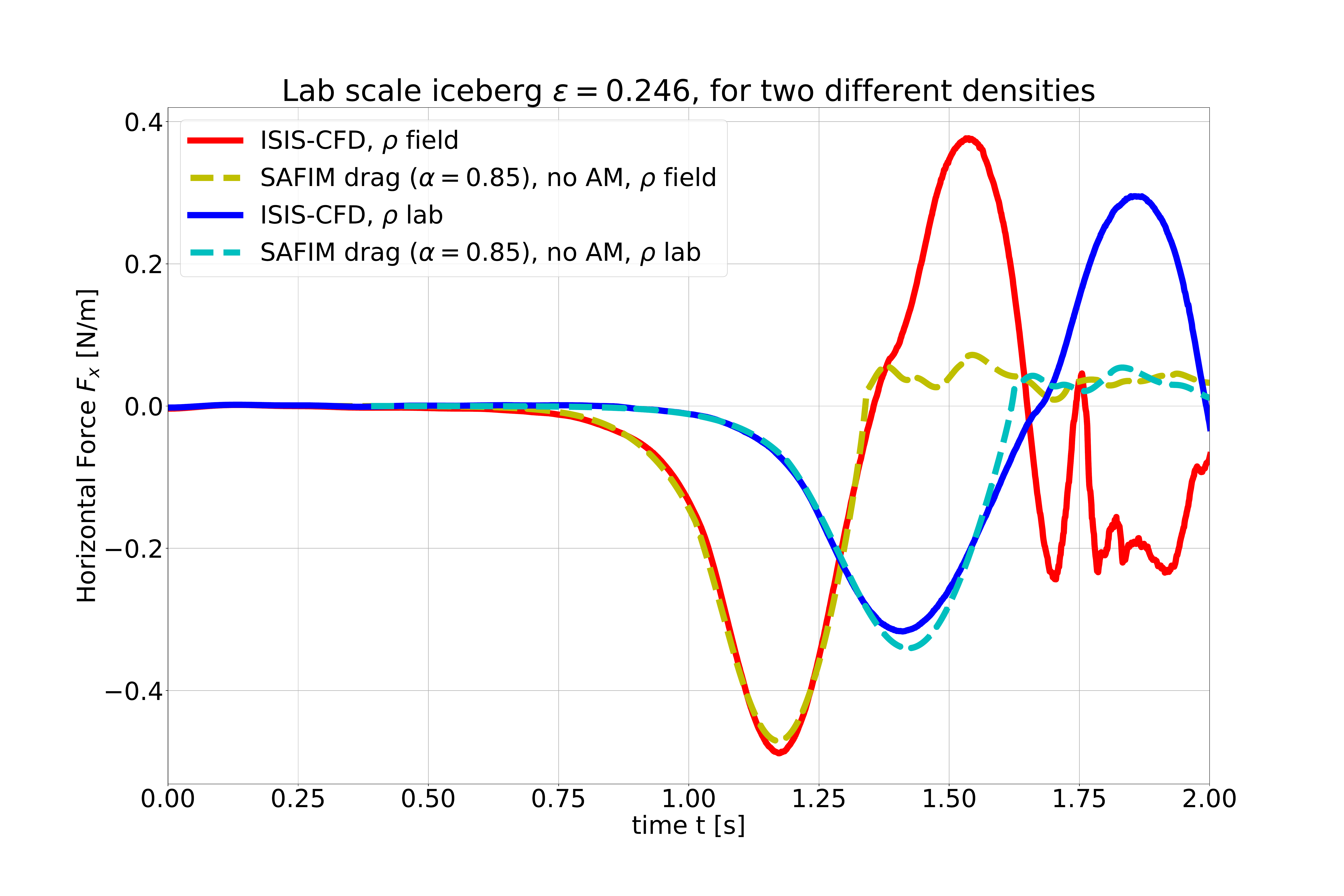}
\caption{Horizontal force acting on a capsizing iceberg ($H=0.103$ m and $\varepsilon=0.246$) computed by ISIS-CFD and SAFIM for two different sets of densities: laboratory densities (blue and cyan curves) $\rho_w=997\;\kilo\gram\per\meter^{3}$ and $\rho_i=920\;\kilo\gram\per\meter^{3}$ and field densities (red and yellow curves) $\rho_w=1025\;\kilo\gram\per\meter^{3}$ and $\rho_i=917\;\kilo\gram\per\meter^{3}$.}
\label{fig:dens}
\end{figure}

In Section~\ref{sec:lab_scale}, we pointed out the similarity between laboratory scale and field scale simulations if the same water and ice densities were used in both.
To obtain the dimensionless variables, we used the time scale $T_H=\sqrt{H/g}$, length scale $H$ and mass scale $m$ (see Table.~\ref{tab:adim}).
However, as shown in Fig.~\ref{fig:dens}, using different densities yields great differences in the horizontal force.
Here, we explain how a simulation of a laboratory-scale iceberg with laboratory densities can be related to a simulation of a field-scale iceberg with field densities.
We use the same approach as in Section~\ref{sec:lab_scale}, with length scale $H$ and mass scale $m$ but we introduce a time scale depending on the densities as proposed by \cite{Tsai2008} (ignoring the factor $2\pi$):

\begin{equation}
T_{\rho,H}=\sqrt{\frac{H \rho_i}{g(\rho_w-\rho_i)}}.
\end{equation}
In Fig.~\ref{fig:adidens}, we plot the dimensionless horizontal force $F'_x = F_x\,T^2/(mH)$ with respect to the dimensionless time $t'=t/T$ for time scale $T=T_H$ and for time scale $T=T_{\rho,H}$ and for three aspect ratios $\varepsilon=0.25$, $\varepsilon=0.375$ and $\varepsilon=0.5$.
For the time scale $T_H$ which does not involve densities, the dimensionless curves differ considerably whereas for $T_{\rho,H}$, which takes the densities into account, the agreement is improved, especially for small aspect ratios.

Using a shift in time, the fit can be improved even further. Therefore, to upscale the laboratory-scale experiments to the field scale, a dimensionless time scale $T_{\rho,H}$ should be used rather than a simple scaling $T_H$.

\begin{figure}[htb!]
\centering
\includegraphics[width=1\linewidth]{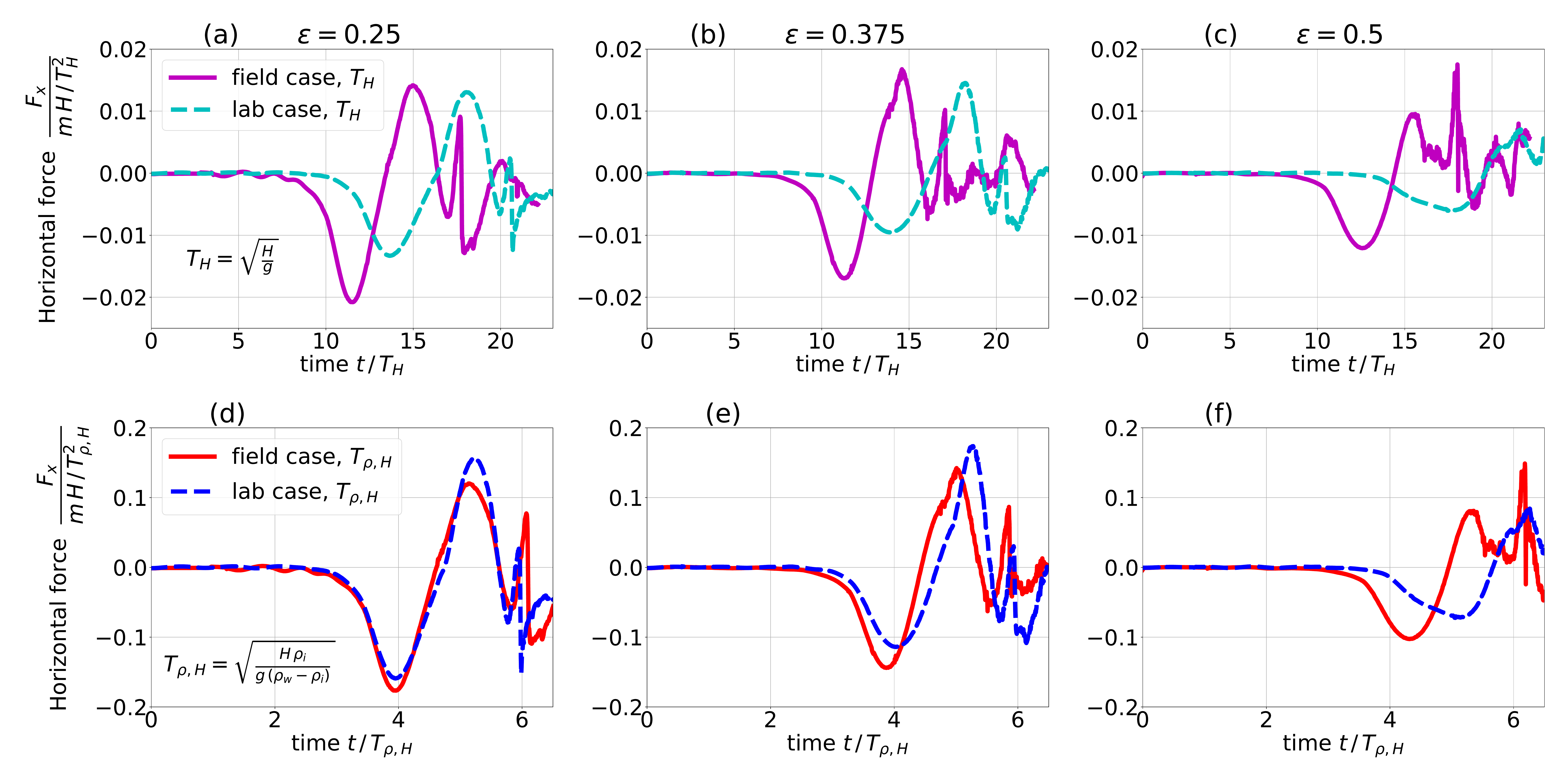}
\caption{Dimensionless horizontal force acting on a capsizing iceberg computed by ISIS-CFD for a field-scale iceberg with field densities $\rho_w=1025\;\kilo\gram\per\meter^{3}$ and $\rho_i=917\;\kilo\gram\per\meter^{3}$ (solid lines) and for a laboratory-scale iceberg with laboratory densities $\rho_w=997\;\kilo\gram\per\meter^{3}$ and $\rho_i=920 \kilo\gram\per\meter^{3}$ (dashed lines). The top row (a, b, c) is for time scale $T_H=\sqrt{H/g}$ and the bottom row (d, e, f) is for time scale $T_{\rho,H}=\sqrt{H/g}\;\sqrt{\rho_i/(\rho_w-\rho_i)}$.
The first, second and third columns correspond to $\varepsilon = 0.25,\; 0.375$ and $0.5$, respectively.
}
\label{fig:adidens}
\end{figure}
As densities have a large impact on capsize dynamics, more realistic water and ice densities, including their spatial heterogeneity, should be considered in future capsize models.
Water density depends on salinity and temperature. For example, in the fjord of the Bowdoin glacier (northwest Greenland), water density may change in the range between $1015\;\kilo\gram\per\meter^{3}$ and $1028\;\kilo\gram\per\meter^{3}$ \cite{Ohashi2019,Sejr2017, Middelbo2018,Holding2019}.
Ice density is more difficult to evaluate as in situ measurements are rare. It depends on the volume fraction of air bubbles, which is for example around $20-30\%$ for firn at $\approx 40\;\meter$ in depth \cite{Herron1980}.
The density of the iceberg may then be heterogeneous and can probably range between $\approx600\;\kilo\gram\per\meter^{3}$ and $\approx930\;\kilo\gram\per\meter^{3}$ 
(the density of pure ice at $-10\degree C$ being about $918\;\kilo\gram\per\meter^{3}$).
With these ranges of ice and water field densities, the factor $\sqrt{\rho_i/(\rho_w-\rho_i)}$ varies between the extreme values $\approx1.18$ and $\approx 3.31$, which corresponds to an even greater spread than in our lab/field comparison ($3.46$ for lab densities, $2.92$ for field values). Therefore consideration of the effect of density and its variability has to be integrated in the inverse problem for iceberg volume estimation based on the seismic signal inversion.

\subsection{3D effects}
\label{sec:3d}

Capsizing icebergs have the following typical dimensions: full-glacier-height $500 \text{ m}\lessapprox H \lessapprox 1000$~m, width in the glacier's flow direction $W\lessapprox 0.75H$ \cite{MacAYEAL2003}, width along the glacier's coast line generally greater than the iceberg's height $H \lessapprox L$, with the upper limit equal to the glacial fjord width.
However, as discussed above, in our modelling we neglect the effect of the third dimension on the dynamics of the capsizing iceberg.
The first argument to support this simplification is that iceberg capsizing in a narrow fjord-like tank (laboratory experiments of \cite{Burton2012}) is very well reproduced with the two-dimensional ISIS-CFD model (Section~\ref{sec:ISIS}).
In the field, icebergs capsize in fjords with much more complex geometries. For example, the fjord may be much wider than the iceberg which would yield a truly three-dimensional motion of the fluid. Real capsizing iceberg should induce vortices on each side of the iceberg which may have an effect on the motion of the iceberg that has not yet been evaluated.

\subsection{Effect of the iceberg geometry}
\label{sec:disgeom}
This study was conducted with the assumption that the icebergs have a perfectly rectangular (parallelepipedic) shape and smooth surface.
However, icebergs in the field have much more complex shapes. The freeboard of an iceberg has irregularities that can range from a scale larger than $100$~m down to a scale less than $0.1$~m \cite{Landy2015}.
The roughness of the submerged part of icebergs is poorly documented because of the difficulty in conducting suitable measurements. In future work, we could estimate the roughness of some well documented icebergs, such as the PII-B-1 tabular iceberg in Northwest Greenland scanned with a Reson 8125 multibeam sonar by \cite{Wagner2014}.
In fluid mechanics modelling, surface features have a great impact on the boundary layer close to the surface and in some cases also on the whole flow \cite{Krogstad1999}.
A sensitivity analysis would be needed to assess the influence of the surface features and surface roughness on the dynamics of capsizing icebergs.

Furthermore, in our simulations, icebergs were initially in hydrostatic equilibrium. In \cite{Sergeant2018}, the effect of hydrostatic imbalance of the iceberg at the initiation of the capsize was assessed by varying the vertical position of the iceberg with respect to the water level. Hydrostatic imbalance results in a different evolution of the contact force with the glacier front and different dominant frequencies of generated seismic waves. This is supported by seismic observations of calving events.

\section{Conclusion}
In this study, we have improved the understanding of free iceberg capsize in open water through fluid-dynamics simulations (ISIS-CFD solver) validated against laboratory experiments \cite{Burton2012}.
In particular, we have shown the complexity of the fluid motion and the dynamics of the iceberg during capsize: vortices around the iceberg during and after capsize, motion of the fluid around the iceberg (velocity of $\approx88$~cm/s for a $H=800$~m high iceberg at a distance H from the iceberg), wave generation, iceberg submergence when reaching the horizontal position and a significant horizontal displacement of the iceberg during capsize. Moreover, we have shown that the non-dimensionalized horizontal force $F'_x=F_x/(\rho_iH^2 \varepsilon g)$ is invariant with the height $H$ of the iceberg. 
The horizontal force acting on the iceberg while its capsize changes its sign after the full capsize. Depending on the iceberg dimensions, this reverse force could be as high and last as long as the one acting during the capsize. Extrapolating these results to iceberg capsize against the glacier terminus would suggest that the force applied by the rotating iceberg at the glacier could be followed by a purely hydrodynamic force of opposite sign, once the contact is lost. This could possibly be compatible with the boxcar force shape assumed by \cite{Olsen2017,Olsen2019} even though the filtering of the contact force itself, with a constant sign, would also lead to a changing sign filtered force as explained by \cite[fig. 7]{Sergeant2018}.
This hypothesis should be however clarified by a full scale CFD analysis including contact and glacier terminus.

We have presented here a Semi-Analytical Floating Iceberg Model (SAFIM) and demonstrated its accuracy for various geometries as well as for different water and ice densities by comparing the results with the direct numerical CFD simulations.
Our simple model is slightly more complex but more accurate than the one used in our previous study~\cite{Sergeant2018}: the new feature is that the drag parameter depends on the iceberg aspect ratio (affine function) to minimize the error with the reference CFD simulations. SAFIM's error is of $5\%$ to $20\%$ (about half the maximum error made with the \cite{Sergeant2018} model) on the horizontal force $F_x$ (without added-masses) during the capsize phase for different aspect ratios. An extension of this model to more complex iceberg shapes and to three dimensions is relatively straightforward.
Different options are offered by SAFIM. For accurate modelling of the amplitude of the fluid forces, SAFIM should be used with drag but without added-masses. 
For accurate modelling of the time of the peak force and the torque, it should be used with a drag force and an added moment of inertia. 
In the global context of estimations of iceberg volume by analysis of seismic signals generated during iceberg capsize in contact with a glacier front, 
based on the discussion on the time-shift in Section~\ref{sec:time}, SAFIM should be used with an optimized drag coefficient and no added-masses. 
However, for today this model has been validated only for the case of capsize of an iceberg in the open ocean. 
Further validation will be conducted for the simulation of the capsize of an iceberg in contact with a glacier. 
In the geophysical context of modelling seismogenic iceberg capsize, further studies would help improve the model accuracy. 
Examples of such studies include (i) modelling of the full glacier / ocean / bedrock / iceberg / ice-m\'{e}lange system, which is computationally very challenging and 
(ii) sensitivity analysis of the iceberg dynamics to the iceberg shape, surface roughness and fjord geometry, which may be also very complex.

\section*{Acknowledgments}
The authors acknowledge funding from ANR (contract ANR-11- BS01-0016 LANDQUAKES), ERC (contract ERC-CG-2013-PE10-617472 SLIDEQUAKES), DGA-MRIS and IPGP - Universit\'{e} de Paris ED560 (STEP'UP), which has made this work possible. The authors acknowledge Justin Burton for providing us with the data from laboratory experiments. The authors are also very grateful to Fran\c cois Charru, Emmanuel de Langre and Evgeniy A. Podolskiy for fruitful discussions, and the reviewers (Jason M. Amundson and Bradley P. Lipovsky) for helpful comments.


\appendix

\section{Dimensional analysis}
\label{sec:simi}
The Vashy-Buckingham - $\pi$ theorem states that the problem can be written with $n-p$ dimensionless ratios obtained by a combination of the $n$ characteristic variables. The integer $p$ is the number of independent physical dimensions in the iceberg capsize system, which is 3 (\textit{time}, \textit{length} and \textit{mass}). The characteristic variables of the system are the dimensions, $H$ and $W$, the densities $\rho_i$ and $\rho_w$, the water viscosity $\mu_w$ and gravity $g$, so $n=6$.

The $n-p=3$ dimensionless ratios are chosen here to be: $$\varepsilon ,\; \cfrac{\rho_w}{\rho_i}, \; \cfrac{\mu_w}{\rho_w H^{3/2}g^{1/2}}$$
The calculation of the horizontal force $F_x(t)$ from the $n=6$ independent characteristic variables of the problem can be written as:
$$F_x=f(H,W,\rho_i,\rho_w,\mu_w,g)$$
The Vashy-Buckingham - $\pi$ theorem states that the problem can be written as:
\begin{equation}
\frac{F_x}{m g} = \mathcal{G}(\varepsilon,\cfrac{\rho_w}{\rho_i},\cfrac{\mu_w}{\rho_w H^{3/2}g^{1/2}})
\label{eq:dim_analysis}
\end{equation}

To estimate the effect of viscosity, we compare the pressure and viscous forces.
The fluid force on the surface of the iceberg calculated by ISIS-CFD is the sum of a friction-induced force (locally tangent to the fluid/solid interface) and a pressure-induced force (normal to this interface).
In the case of an iceberg with aspect ratio $\varepsilon = 0.25$, the friction force is found to be $\approx 300$ times smaller than the pressure force for the field-scale case ($H=800~\meter$) and $\approx 10$ times smaller for the laboratory case ($H=0.103~\meter$) as illustrated in Fig.~\ref{fig:fxpv}.
Therefore, viscous effects can be reasonably neglected in both scales.
This leads to the following approximation for equations (\ref{eq:dim_analysis}):
\begin{equation}
\frac{F_x}{m g} \approx \mathcal{G}(\varepsilon,\cfrac{\rho_w}{\rho_i}) \;,
\end{equation}
\textit{i.e.} for similar initial conditions and boundary conditions, the evolution with time of the dimensionless force $F'_x=F_x / (m g)$ only depends on the aspect ratio $\varepsilon$ and the density ratio $\rho_w / \rho_i$. However, the function $f$ remains unknown and is investigated in Section~\ref{sec:lab_scale} and Section~\ref{sec:density}.

\begin{figure}[htb!]
	\centering
	\includegraphics[width=1\linewidth]{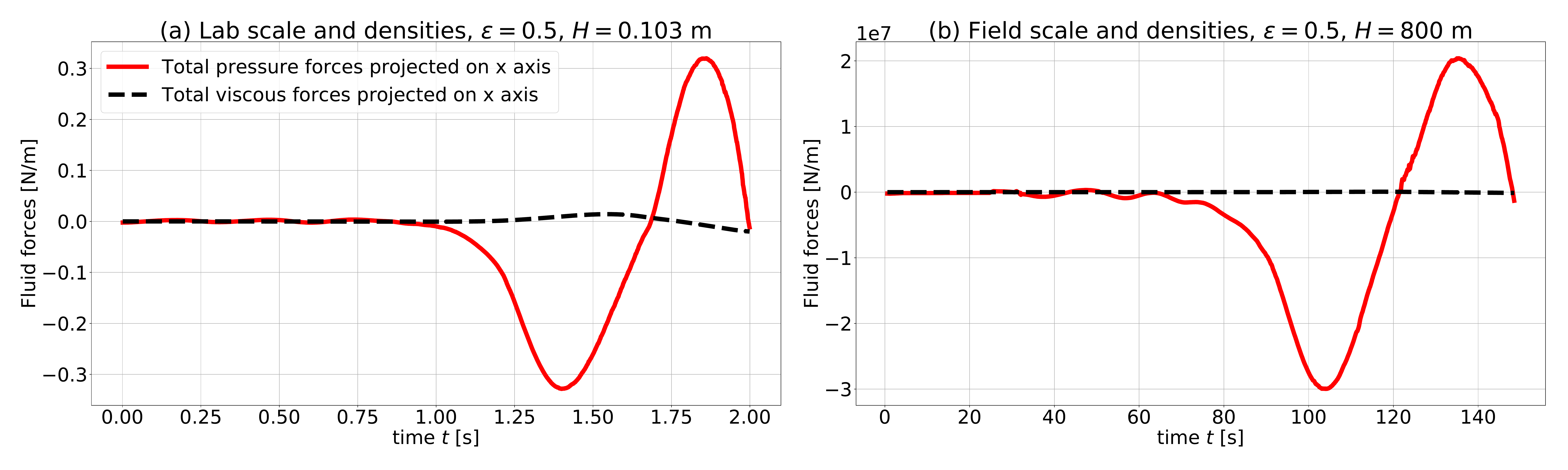}
	\caption{Viscous forces and pressure forces for (a) a laboratory-scale and (b) field-scale iceberg capsize, computed by ISIS-CFD.}
	\label{fig:fxpv}
\end{figure}

\section{SAFIM forces and torques}
\label{sec:expressions}

The integrated expressions for hydrostatic and drag forces and the associated torques are given below for SAFIM and for a rectangular iceberg as in Fig.~\ref{fig:schema_iceberg_dz0}.
All these expressions are implemented in the Python code available online at \cite{vladislav_yastrebov_2019}.

The effect of the hydrostatic pressure is given by the following integral:
$\vec F_s = -\rho_w g\int\limits_{\Gamma_s} z \vec n\,d\Gamma$,
where $\vec n$ is the outward surface normal and $\Gamma_s$ is the submerged part of the iceberg.
The torque induced by this pressure distribution with respect to the centre of gravity $G$ at position $\vec r_G$ is given by:
$\vec M_s = -\rho_w g \int\limits_{\Gamma_s}(\vec r - \vec r_G)\wedge \vec n\,d\Gamma$

The drag force is given by Eq.~\ref{eqn:DragSAFIM1}, and the drag torque with respect to G by Eq.~\ref{eqn:DragSAFIM2}.
The calculation of the integral of the pressure drag is split into integration over all submerged or partly submerged sides of the rectangular iceberg. Consider a partly submerged side $S=AB$ and let us assume that corner $A$ is a submerged corner and $B$ is a corner outside the water. Then the velocity $\vec v_M$ of a point $M \in [AB]$ is:

\begin{equation}
\vec v_M= \vec v_G +  \dot\theta \vec e_y \wedge (\vec r_M - \vec r_G),
\end{equation}

where $\vec v_G$ is the velocity of G, $\vec r_M = \vec r_A + \xi  (\vec r_B - \vec r_A)$ with $\xi \in [0,\xi_i]$ and $\xi_i$ defining the intersection between$[AB]$ and the water surface.
Thus, for the side $AB$, the contribution of the drag force is given by
$$
F_d^{AB} =  \frac12 \alpha \rho_w \vec n \|\vec r_B - \vec r_A\| \int\limits_{0}^{\xi_i}  |v_n|^2 \mbox{sign}(v_n) \,d\xi,
$$
where 
$v_n = v_M \cdot\vec n.$
For the case of a totally submerged side, $\xi_i=1$. For the case of a side totally outside the water, the contribution to the drag force is zero.

\subsection{Horizontal motion of capsizing iceberg}

With the formulation of the drag force given above, we can reproduce the horizontal motion of a freely capsizing iceberg, which is observed experimentally and reproduced with the accurate ISIS-CFD simulations.
Obtaining a closed form solution of SAFIM equations Eqs.~\eqref{eqn:alleq1},~\eqref{eqn:alle2} and~\eqref{eqn:alleq3} is out of reach.
We wish to give here some intuitive explanation of the horizontal motion of the iceberg.
The resultant of the buoyancy and gravity forces moves the iceberg upwards and makes it rotate: these two effects initiate the vertical and rotational motion of the iceberg. The induced velocity produces a force with a non-zero horizontal component.

We now explain why these two initial motions -upwards and rotation-, together generate a horizontal drag force, in the framework of SAFIM.
We draw the velocity $\vec{v}$ (triple red arrow) of several points on the surface of the iceberg and its normal component $v_n \boldsymbol{n}$ (dashed red arrow).
In SAFIM, the elementary drag force $d\vec F_d$ (solid black arrow) is collinear with $\vec n$ and opposes the normal velocity $v_n\vec n$. 
The projection of the elementary drag forces on the horizontal axis $d\vec F_{d_x}=(d\vec F_d \cdot \vec e_x ) \vec e_x$ 
is shown by a dashed green arrow if it is leftward and dashed blue arrow if it is rightward.
The integral of these horizontal elemental forces results in the global horizontal force $\vec F_{d}\cdot \vec e_x$.

For the case of upward motion Fig.~\ref{fig:dragexp}(a), the vertical local velocity is constant along the iceberg surface.
Therefore, the horizontal drag force on the two long sides of the iceberg is leftward elementary force whereas on the small submerged side $CD$, it is rightward but with a smaller amplitude. The iceberg thus moves to the left.

For the case of rotational motion around $G$, with no motion of $G$ (Fig.~\ref{fig:dragexp}(b)), the velocity increases with the distance to the centre of rotation. The further away the point is, the more it contributes to the drag force.
Two points located at the same distance from $G$, but with opposite normal velocity $v_n \vec  n$ (i.e. one point on the blue line and one point on the green line) have the same absolute contribution to the drag force but in opposite directions. Thus the drag force on the part of the surface coloured by solid green lines compensates the drag force on the part coloured blue.
The remaining part of the surface coloured by dashed green lines at the iceberg bottom, induces a leftward total horizontal force.
Therefore, by superposing vertical and rotation motion, we obtain a net drag force in the direction of the initial tilt of the iceberg, here to the left.

\begin{figure}[htb!]
\centering
\begin{minipage}{1\linewidth}
\centering
\includegraphics[width=1\linewidth]{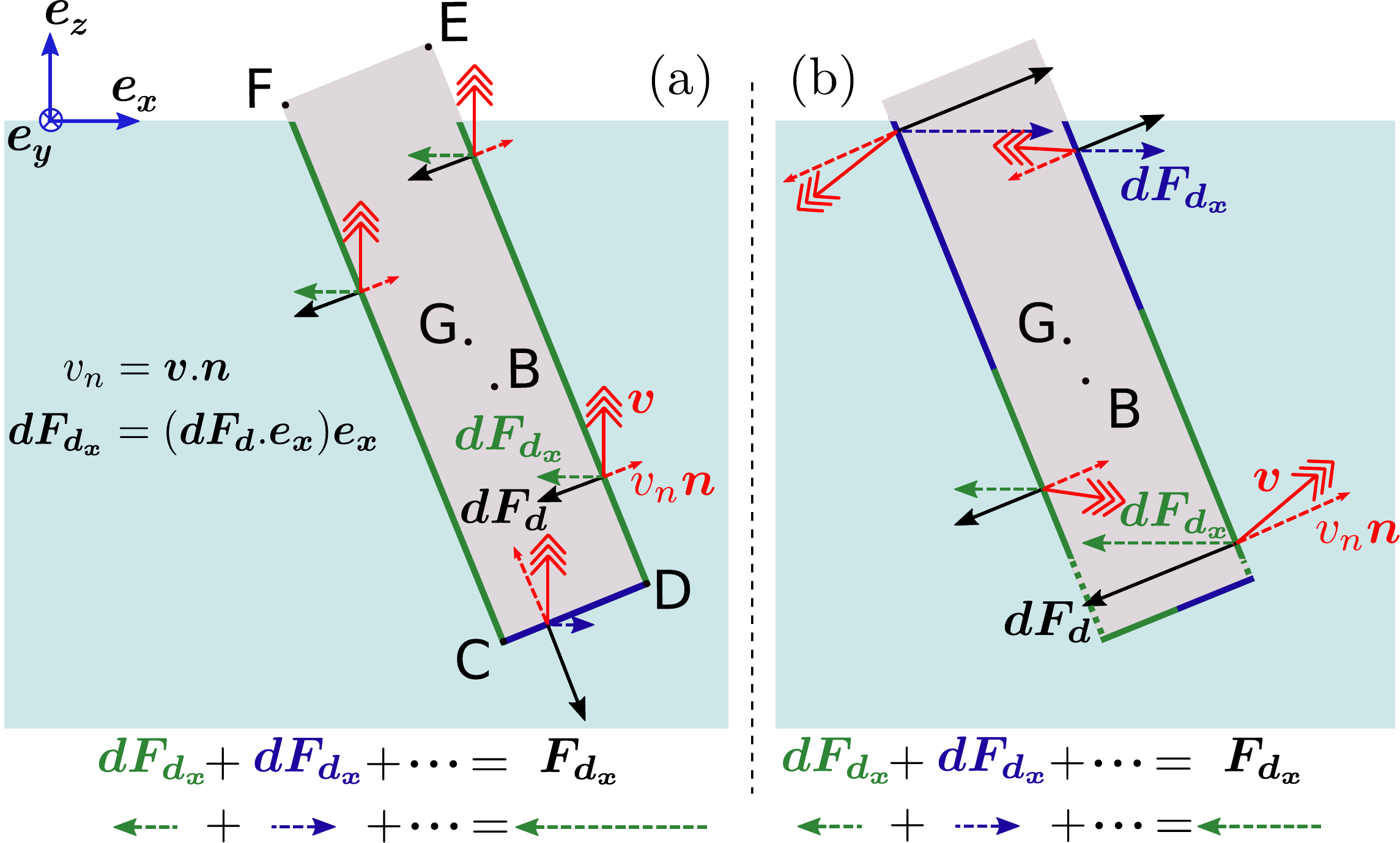}
\end{minipage}
\caption{Schematic explanation of the horizontal force induced by the formulation of the drag force in SAFIM (a) vertical upward motion of iceberg, (b) rotation of iceberg.\label{fig:dragexp}}
\end{figure}

\subsection{Simplified added-masses}
\label{sec:expressionsAM}
The calculation of the simplified added-masses given in equation~\eqref{eqn:AMSAFIM} and~(8),
requires the calculation of the effective height $H_{\mathrm{eff}}$ and effective width $W_{\mathrm{eff}}$ of the submarine part of the iceberg (see Fig.~\ref{fig:schema_iceberg_dz0}).
To calculate them, we use the positions of the four corners: $C$, $D$, $E$, $F$ (Fig.~\ref{fig:dragexp}).
The coordinates of a corner $P \in \{C,D,E,F\}$ have the general expression :
$$x_{P}=z_G+\delta_1^P\;\frac{H}{2}\cos(\theta)+\delta_2^P\;\frac{L}{2}\sin(\theta)$$
$$z_{P}=z_G+\delta_3^P\;\frac{H}{2}\cos(\theta)+\delta_4^P\;\frac{L}{2}\sin(\theta)$$
with ($\delta_1^P$,$\delta_2^P$,$\delta_3^P$,$\delta_4^P$) defined as follows for the four corners
\begin{equation*}
\begin{split}
&\{\delta_1^C,\delta_2^C,\delta_3^C,\delta_4^C\} = \{1,-1,-1,-1\}, \quad \{\delta_1^D,\delta_2^D,\delta_3^D,\delta_4^D\} = \{1,1,-1,1\},\\
&\{\delta_1^E,\delta_2^E,\delta_3^E,\delta_4^E\} = \{-1,1,1,-1\}, \quad \{\delta_1^F,\delta_2^F,\delta_3^F,\delta_4^F\} = \{-1,-1,1,1\}.\\
\end{split}                                                                                                                                                                                                                                                
\end{equation*}

The effective height can be calculated with the following expressions :
$$H_{\mathrm{eff}}=\mathrm{max}\left( (z_w-z_C)\;;\;(z_w-z_D)\;;\;(z_w-z_E)\;;\;(z_w-z_F) \right) $$
where $z_w$ is the water level.

The effective width, defined as the distance between the leftmost and the rightmost points of the submerged part of the iceberg, is calculated similarly, but after checking which are the submerged corners and the geometrical intersection between the water surface and the iceberg sides.

\section{Optimal parameters}
\label{sec:tabparam}
We summarize in Table~\ref{tab:err} the errors of SAFIM computed with respect to the ISIS-CFD results for a quadratic drag model and the three options for added-masses (no added-masses, simplified or computed added-masses).
These errors correspond to the minimal possible errors obtained by the minimization procedure. The step used for the drag coefficient $\alpha$ was $0.05$ and the step for the added-masses factors was $0.25$.

\begin{table}
\begin{center}
\begin{tabular}{  c  c  | c  c | c  c  | c  c  c  c  c  c  c  c  c }
     \multicolumn{2}{c|}{Parameters} & \multicolumn{2}{|c|}{no AM}& \multicolumn{2}{|c|}{computed $AM$}& \multicolumn{5}{|c}{simplified $AM$} $\displaystyle\vphantom{\frac{A}{A}}$\\
       \cline{1-11}
    $\displaystyle\vphantom{\frac{A}{A}}$ $\varepsilon$ & $\theta_0$ [\textsuperscript{o}] & Error $E_1$ & $\alpha$ & Error $E_2$ & $\alpha$ & Error $E_2$ &$\alpha$&$C_{x}$&$C_{z}$&$C_{I}$ \\
       \cline{1-11}
    $\displaystyle\vphantom{\frac{A}{A}}$0.246 & 0.5 & $5.2\%$ & $0.85$ & $36.6\%$ & $3$ & $10.0\%$ & $1.1$ & $0.$ & $0.$ & $0.75$\\
    0.374 & 0.5 & $9.6\%$ & $1.55$ & $34.1\%$ & $1.8$ & $21.3\%$ & $1.4$ & $0.$ & $0.$ & $0.75$\\
    0.496 & 0.5 & $20.1\%$ & $2.9$ & $40.0\%$ & $1.9$ & $23.0\%$ & $3.0$ & $0.$ & $0.$ & $0.5$\\
    0.639 & 15 &$24.7\%$ & $4.0$ & $47.4\%$ & $2.6$ & $26.2\%$ & $4.2$ & $0.$ & $0.$ & $0.25$\\
  \end{tabular}
\end{center}
\caption{First two columns : geometrical characteristics and initial conditions of the studied icebergs. Laboratory-scale iceberg simulations have height $H=0.1$ m and field-scale iceberg simulations have height $H=800$~m. The density of the water is $\rho_w=1025$~kg~m\textsuperscript{-3} and the density of the ice is $\rho_i=917$~kg~\textsuperscript{-3}. Next columns : parameters minimizing the error on $F_x$ and the corresponding error for SAFIM without added-masses, SAFIM with computed added-masses and SAFIM with simplified added-masses.}
\label{tab:err}
\end{table}

\section{Comparison of SAFIM with model in Sergeant et al. (2018,2019)}
\label{sec:alpha1}
We compare SAFIM model with optimised $\alpha$-factor, SAFIM model with $\alpha=1$ (as used in \cite{Sergeant2018, Sergeant2019}) and ISIS-CFD in Fig.~\ref{fig:alpha1}. The optimisation of the drag coefficient $\alpha$ improves the horizontal force $F'_x$ and torque $M'_{\theta}$, in particular for the three biggest aspect ratios (Figs.~\ref{fig:alpha1}~(c-h)).
\begin{figure}[htb!]
\centering
\begin{minipage}{1\linewidth}
\centering
\includegraphics[width=1\linewidth]{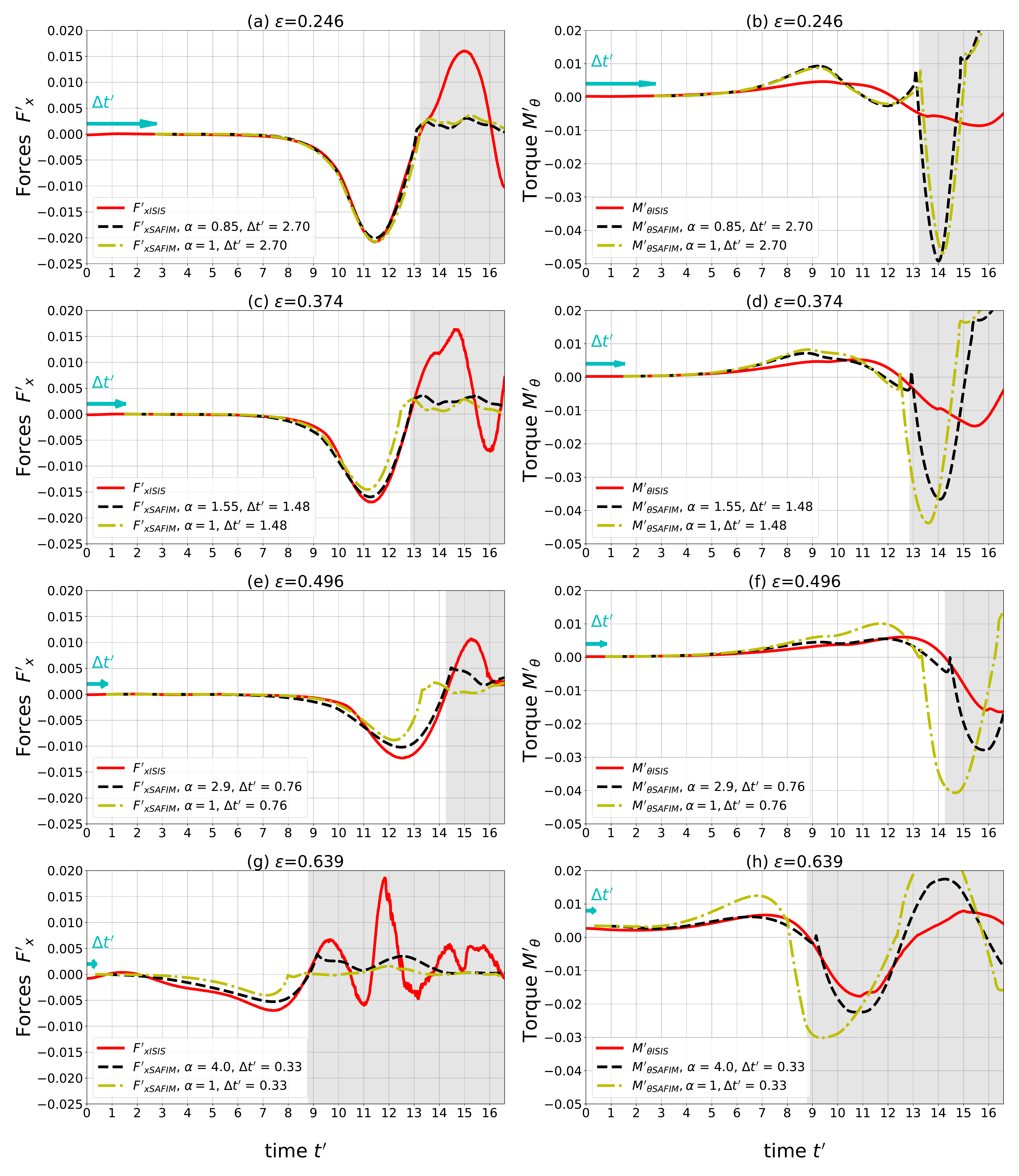}
\end{minipage}
\caption{Capsize simulations for SAFIM with optimised drag without added-masses and with time shits (black lines), for SAFIM with drag coefficient $\alpha=1$ without added-masses and with time shits (yellow lines), and for ISIS-CFD (red lines): evolution of the dimensionless total horizontal force $F'_x$ on the iceberg (a, c, d, e), and torque $M'_{\theta}$ (b, d, f, g). Results are given for icebergs with $\varepsilon=0.246$ (a, b), $\varepsilon=0.374$ (c, d), $\varepsilon=0.496$ (e, f) and $\varepsilon=0.639$ (g, h). SAFIM curves were shifted (blue arrow) by the dimensionless time $\Delta t'= \Delta t \, \sqrt{g/H}$. The SAFIM $\alpha$-factor and time $\Delta t'$ are indicated in the legends.\label{fig:alpha1}}
\end{figure}
%

\end{document}